\documentclass[a4paper,10pt]{article}
\usepackage[utf8]{inputenc}
\usepackage{graphicx}
\usepackage{amsmath}
\usepackage[title,titletoc]{appendix}
\usepackage{array}
\usepackage{longtable}
\usepackage{capt-of}
\usepackage{supertabular}
\usepackage{floatrow}
\usepackage{amssymb}
\usepackage{xcolor}
\usepackage{amssymb}
\usepackage{multicol}
\usepackage{authblk}
\usepackage{geometry} 
\newcolumntype{C}[1]{>{\centering}m{#1}}

\newcommand{\mycirc}[1][black]{\Large\textcolor{#1}{\ensuremath\bullet}}

\newcount\n
\def\tablebody{}
\makeatletter
\loop\ifnum\n<100
        \advance\n by1
        \protected@edef\tablebody{\tablebody
                \textbf{\number\n.}& shortText
                \tabularnewline
        }
\repeat

\makeatletter
\let\mcnewpage=\newpage
\newcommand{\TrickSupertabularIntoMulticols}{%
  \renewcommand\newpage{%
    \if@firstcolumn
      \hrule width\linewidth height0pt
      \columnbreak
    \else
      \mcnewpage
    \fi
  }%
}
\makeatother

\title{Multiplexity versus correlation: the role of local constraints in real multiplexes}

\author[*]{V. Gemmetto}
\author[*]{D. Garlaschelli}
\affil[*]{Instituut-Lorentz for Theoretical Physics, Leiden Institute of Physics, \newline University of Leiden, The Netherlands}

\date{}

\begin{document}

%%%%%%%%%%%
%\twocolumn[
  %\begin{@twocolumnfalse}
  
    \maketitle
    
\begin{abstract}
Several real-world systems can be suitably represented as multi-layer complex networks, i.e. in terms of a superposition of various graphs, 
each related to a different mode of connection between nodes. 
Hence, the definition of proper mathematical quantities aiming at capturing the added level of complexity of those systems is required. 
Various attempts have been made in order to measure the empirical dependencies between the layers of a multiplex, for both binary and weighted networks.
In the simplest case, such dependencies are measured via correlation-based metrics, a procedure that we show to be equivalent to the use of completely homogeneous benchmarks specifying only global constraints, such as the total number of links in each layer. 
However, these approaches do not take into account the heterogeneity in the degree 
and strength distributions, which instead turns out to be a fundamental feature of real-world multiplexes. In this work, we compare the observed dependencies
between layers with the expected values obtained from
reference models that appropriately control for the observed heterogeneity in the degree and strength distributions.
This results in the introduction of novel and improved multiplexity measures that we test on different datasets, i.e. the International Trade Network (ITN) and the European Airport Network (EAN). 
Our findings confirm that the use of homogeneous benchmarks can lead to misleading results, and highlight the important role played by the distribution of hubs across layers. 
In the EAN, where different layers are characterized by different hubs, the multiplexity is practically absent, irrespective of the null model considered.
By contrast, in the ITN, where the same nodes tend to be hubs in all layers, the seemingly strong inter-layer correlations are severely reduced once the local constraints are controlled for, meaning that most dependencies are actually encoded in the correlated degree (or strength) sequence of the multiplex. 
This shows that a strong similarity among all layers of a multiplex is not necessarily a genuine signature of multiplexity.
\end{abstract}
Several real-world systems can be represented as multi-layer complex networks, i.e. in terms of a superposition of various graphs, 
each related to a different mode of connection between nodes. 
Hence, the definition of proper mathematical quantities aiming at capturing the level of complexity of those systems is required. 
Various attempts have been made to measure the empirical dependencies between the layers of a multiplex, for both binary and weighted networks.
In the simplest case, such dependencies are measured via correlation-based metrics, a procedure that we show to be equivalent to the use of completely homogeneous benchmarks specifying only global constraints, such as the total number of links in each layer. 
However, these approaches do not take into account the heterogeneity in the degree 
and strength distributions, which instead turns out to be a fundamental feature of real-world multiplexes. In this work, we compare the observed dependencies
between layers with the expected values obtained from
reference models that appropriately control for the observed heterogeneity in the degree and strength distributions.
This results in the introduction of novel and improved multiplexity measures that we test on different datasets, i.e. the International Trade Network (ITN) and the European Airport Network (EAN). 
Our findings confirm that the use of homogeneous benchmarks can lead to misleading results, and highlight the important role played by the distribution of hubs across layers. 
In the EAN, where different layers are characterized by different hubs, the multiplexity is practically absent, irrespective of the null model considered.
By contrast, in the ITN, where the same nodes tend to be hubs in all layers, the seemingly strong inter-layer correlations are severely reduced once the local constraints are controlled for, meaning that most dependencies are actually encoded in the correlated degree (or strength) sequence of the multiplex. 
This shows that a strong similarity among all layers of a multiplex is not necessarily a genuine signature of multiplexity.
%\end{@twocolumnfalse}
%]

\section{Introduction}

The study of networks has been pursued in order to suitably represent biological, economic and social systems, exploiting the possibility to analyze such
structures as a set of units connected by edges symbolizing interactions among those elements~\cite{Barabasi, Newman1, Watts, Fortunato}.

However, this assumption may actually lead to an oversimplification;
indeed, several systems are composed by units connected by multiple interactions.
In such systems, the same set of nodes is joined by various types of links, each of those representing a different mode of connection~\cite{Wasserman}. 
For instance, a given set of individuals may be connected by multiple on-line social networks, therefore exchanging
information with different neighbors depending on the layer~\cite{Szell}; moreover,
the system of alternative means of transportation between places in a city (bus, tram, metro, etc.) may be suitably represented 
as the superposition of various interdependent networks~\cite{Zou, Kurant}.
The simplest way to analyse such systems is the aggregation of the various levels in a single network, but it turns out that 
such a simplification may discard crucial information about the real topology of the network and therefore about possible dynamical processes acting
on the system~\cite{Buldyrev}. For instance, such an aggregation may result in a loss of information about the distribution of the hubs across layers,
which is instead crucial for the control of several processes arising on an interdependent network~\cite{Radicchi}.
Then, in order to solve such an issue, in the last few years the study of multi-layer networks has been pursued. 
In this context, new quantities aiming at mathematically analyzing multi-level networks have been provided~\cite{Szell, Cardillo, Morris, Battiston}; 
furthermore, models of growth~\cite{Nicosia1, Kim, Nicosia2} and  
dynamical processes occurring on multiplexes, such as epidemic spreading~\cite{Saumell}, diffusion~\cite{Gomez}, cooperation~\cite{GomezGar} 
and information spreading~\cite{Estrada} have been designed.

In this work, we follow the path towards the definition of measures that can be applied to multi-level networks, in order to characterize crucial structural properties 
of these systems, in particular focusing on the analysis of the correlations between layers.
Indeed, correlations, embodied by layers overlap, represent an important feature for many real-world multiplexes; for instance, going back again to the virtual social network
case, we should expect a high overlap between the various layers, since individuals are likely to be connected to their friends in most of the on-line social networks they use, 
thus increasing the overlay between layers.
Hence, proper measures of correlation in multi-level systems are needed.
However, a comparison between the observed correlation and some notion of expected correlation is required. 
In this context, we exploit the concept of multiplex ensembles~\cite{Newman2, Park1, Bianconi}, aiming at the definition of suitable null models for multi-layer complex networks, 
in order to compare the observed overlap between layers with the expected overlap we would find due to a random superposition.    

Various efforts have already been made about the study of correlations in multi-level networks~\cite{Barigozzi,Nicosia3, Lee}, but the comparison of the observed results with the expected ones
has generally been based on a - sometimes implicit - assumption: the benchmark was a completely homogeneous graph. 
In particular, here we show that correlation-based measures of inter-layer dependency (of the type used e.g. in ref.\cite{Barigozzi}) build on an implicit assumption of homogeneity, which in the unweighted case is equivalent to the choice of
the random graph as null model. 
Similarly, for weighted networks, the chosen benchmark was equivalent to the weighted random graph, where the weight
distribution is independent from the considered pair of nodes~\cite{Garlaschelli1}.

However, this assumption of uniformity in the probability distributions strongly contrasts with the observed findings in real-world complex systems. Indeed,  
one of the most well-known features of complex networks is their heterogeneity~\cite{Barrat}, both in the degree distribution and in the weight
distribution; it is therefore crucial to take this aspect into account when proper null models for graphs are designed.
Here we use, as benchmarks, models taking into account the heterogeneity observed in the degrees (for unweighted networks) or in the strengths (for weighted graphs), 
showing that such a refinement can completely change the final findings and lead to a deeper understanding of the actual 
correlations observed between layers of a real-world multiplex.

We introduce a new measure of multiplexity designed to quantify the overlap between layers of a multi-level complex
network. Furthermore, we derive the expression of the expected value of such a quantity, both in the binary and in the weighted case, for randomized networks, by enforcing different
constraints.
We then apply our measures to two different real-world multiplexes, namely the International Trade Network and the European Airport Network, showing that the analysis
of the correlations between layers can actually make some important structural features of these systems explicit.

Indeed, while the former shows significant correlations between layers (i.e., traded commodities), in the latter almost no overlap can generally be detected, thus clearly 
defining two opposite classes of multiplexes based on the observed correlations. Furthermore, we will link such a behaviour with the distribution of the hubs across layers, 
hence providing a straightforward explanation to the observed findings.

\section{Methods}

\subsection{Uncorrelated null models for multi-layer networks}

As in previous studies~\cite{Bianconi}, we define the multiplex $ \overrightarrow{G} = (G_1,G_2,\ldots,G_M) $ as the superposition of $ M $ layers $ G_k $ ($ k = 1,2,\ldots,M $), each of them represented by a
(possibly weighted) network sharing the same set of $ N $ nodes 
with the other ones, although we do not require that all the vertices are active in each layer. 
Therefore, multiplex ensembles can be defined by associating a probability $ P(\overrightarrow{G}) $ to each multi-network, so that the entropy $ S $ of the ensemble is given by:
\begin{eqnarray}
S = -\sum_{\overrightarrow{G}}^{} P(\overrightarrow{G}) \ln P(\overrightarrow{G})
\end{eqnarray}
It is then possible to design null models for multi-level networks by maximizing such an entropy after the enforcement of proper constraints. In this context,
previous works~\cite{Bianconi, Halu, Menichetti} introduced the concepts of correlated and uncorrelated multiplex ensembles, based on the possibility to 
introduce correlations between
layers within the null models. In particular, for an uncorrelated ensemble the probability of a given multiplex can be factorize into the probabilities
of each single-layer network $ G_k $ belonging to that multiplex, as the links in any two layers $ \alpha $ and $ \beta $ are uncorrelated; thus, it is given by:
\begin{eqnarray}
P(\overrightarrow{G}) = \prod_{k = 1}^{M} P_k(G_k)
\label{prob_multiplex}
\end{eqnarray}
Instead, if we want to take into account correlations between layers, the previous relation (\ref{prob_multiplex}) does not hold.

Since our purpose is precisely that of measuring such correlations, we will consider the former type of ensemble, in order to define a null model for the 
real system
so that it is possible to compare the observed correlations with reference models where the overlap between layers is actually randomized and, at the same time, 
important properties of the real network are preserved. 

In this perspective, therefore, the definition of proper null models for the considered multiplex reduces to the definition of an indipendent
null model for any layer of the system. 
In order to do this, we take advantage of the concept of canonical network ensemble, or exponential random graph~\cite{Park2}, i.e. the randomized family of graphs satisfying a set of constraints
on average. In this context the resulting randomized graph preserves only part of the topology of the considered real-world network and is entirely
random otherwise, thus it can be employed as a proper reference model.

However, fitting such previously defined models~\cite{Bianconi, Halu, Menichetti} to real datasets is hard, since it is usually computationally demanding as it
requires the generation of many randomized networks whose properties of interest have to be measured. 
%%% fast method
In this perspective, we exploit a fast and completely analytical Maximum Entropy method, based on the maximization of the likelihood 
function~\cite{Garlaschelli0, Squartini1, Squartini6}, which provides the exact 
probabilities of occurrence of random graphs with the same average constraints as the real network. From such probabilities it is then possible to compute the expectation values of 
the properties we are interested in, such as the average link probability or the average weight associated to the link established between any two nodes.
This procedure is general enough to be applied to any network, including the denser ones, and does not require the sampling of the configuration space in order to 
compute average values of the quantities of interest.
While the adoption of such a method is not strictly required when dealing with global constraints like the total number of links observed in a network, it becomes 
crucial when facing the problem of enforcing local constraints such as the degree sequence or the strength sequence. 

Before introducing our measures of multiplexity, we make an important preliminary observation. Simple measures of inter-layer dependency are based on correlation metrics, which in turn rely on an assumption of uniformity, such assumption being ultimately equivalent to the choice of a uniform random graph as a null model.
We illustrate this result in detail in the Appendix.
As such, these naive measures completely disregard the observed structural heterogeneity of the multiplex. For instance, a correlation measure introduced in~\cite{Barigozzi} - for 
both binary and weighted networks - implicitly build on such a homogeneous assumption, therefore discarding most of the information encoded in the considered real system (see Appendix).

%%% homo
Indeed, so far the most widely used graph null model has been represented by the random graph (RG)~\cite{Park2}, which enforces on average as constraint the expected number of links in the network.
Such model, therefore, provides a unique expected probability $ p_{\alpha} $ that a link between any two nodes is established in layer $ \alpha $: however, as we said, such a 
reference model completely discards any kind of heterogeneity in the degree distributions of the layers, resulting in graphs where each node has on average the same number 
of connections, inconsistently with the observed real networks.
Thus, the probability of connection between any two nodes in layer $ \alpha $ is uniformly given by:
\begin{eqnarray}
p_{\alpha} = \frac{L_{TOT}^{\alpha}}{N(N - 1)/2}
\label{p_alfa_bin}
\end{eqnarray}
where $ L_{TOT}^{\alpha} $ is the total number of links actually observed in layer $ \alpha $:
\begin{eqnarray}
L_{TOT}^{\alpha} = \sum_{i < j}^{} a_{ij}^{\alpha}
\end{eqnarray}
and $ a_{ij}^{\alpha} = 0,1 $ depending on the presence of the link between nodes $ i $ and $ j $ in layer $ \alpha $.

Similar considerations apply to weighted networks and the related weighted random graph (WRG)~\cite{Garlaschelli1}, i.e. the straightforward extension of the previous random graph 
to weighted systems; in such a null model, the probability of having a link of weight $ w $ between two nodes $ i $ and $ j $ is independent from the choice of the nodes, and it 
is given by the following geometric distribution:
\begin{eqnarray}
P(w^{\alpha}) = {p_{\alpha}}^{w} (1 - p_{\alpha})
\end{eqnarray}
where the Maximum Likelihood method shows that the optimal value of the parameter $ p_{\alpha} $ is given by:
\begin{eqnarray}
p_{\alpha} = \frac{2 W_{TOT}^{\alpha}}{N(N - 1) + 2 W_{TOT}^{\alpha}}
\label{p_alfa_wei}
\end{eqnarray}
with $ W_{TOT}^{\alpha} $ defined as the total weight observed in layer $ \alpha $ ($ w_{ij}^{\alpha} $ is the weight associated to the link between
nodes $ i $ and $ j $ in the same layer):
\begin{eqnarray}
W_{TOT}^{\alpha} = \sum_{i < j}^{} w_{ij}^{\alpha}
\end{eqnarray}
Similarly to the corresponding binary random graph, also this kind of null models discards the crucial presence of nodes characterized by higher strengths (that is, by
a higher sum of the weights associated to links incident on that node).

%%% hetero 
To take into account the heterogeneity of the real-world networks within the null models, in the unweighted case we consider the (binary) configuration model 
(BCM)~\cite{Maslov}, i.e. the ensemble 
of networks satisfying on average a given degree sequence. Since we make use of the canonical ensembles, it is possible to obtain from the Maximum Likelihood method
each probability $ p_{ij}^{\alpha} $ that nodes $ i $ and $ j $ are connected in layer $ \alpha $ (notice that such value $ p_{ij}^{\alpha} $ is basically the 
expectation value of $ a_{ij}^{\alpha} $ under the chosen configuration model). Similarly, for weighted graphs the weighted configuration model (WCM)~\cite{Serrano} 
can be defined: here, the enforced constraint is represented by the strength sequence as observed in the real-world network. In this view, the likelihood 
maximization provides the expectation value of each weight $ w_{ij}^{\alpha} $ for any pair of nodes $ i $ and $ j $ as supplied by the weighted 
configuration model. 
It is worth noticing that enforcing the degree sequence (respectively, the strength sequence in the weighted case) automatically leads to the design of a null model where
also the total number of links (respectively, the total weight) of the network is preserved.
In the Appendix, we will provide equations generalizing equations (\ref{p_alfa_bin}) and (\ref{p_alfa_wei}), whose solution allows then to derive the analytical 
expression of the expected link probability $ p_{ij}^{\alpha} $ and, in the weighted case, the expected link weight $ w_{ij}^{\alpha} $. In order to do this, we make use of a set 
of $ N $ auxiliary variables $ x_i^{\alpha} $ for any layer $ \alpha $, which are proportional to the probability of establishing a link between a given node $ i $ and any other 
node (or, respectively for the weighted case, establishing a link characterized by a given weight), being therefore directly informative on the expected probabilities 
$ p_{ij}^{\alpha} $ (or, respectively, the expected weights $ w_{ij}^{\alpha} $).

The previous null models will therefore represent our benchmarks for the analysis of the significant overlaps between layers of a real-world multiplex.
In order to measure the correlation between pairs of layers we introduce the so-called multiplexity, both for binary and weighted multi-layer networks. 
We will then apply such definition to a couple of different real-world systems, showing that in general homogeneous null models as the random graph cannot 
exhibit correlations between layers similar to what we measure in the observed ones.
%%%%%%%%%%%%%%%%%%%%%%%%%%%%%

\subsection{Binary multiplexity}
% binary
When the unweighted networks are considered we define the binary multiplexity between any two layers $ \alpha $ and $ \beta $ as:

\begin{eqnarray}
m^{\alpha, \beta}_{bin} = \frac{2 \sum_{i < j}^{} \min \{ a_{ij}^{\alpha}, a_{ij}^{\beta}\}}{L_{TOT}^{\alpha} + L_{TOT}^{\beta}}
\label{m_bin}
\end{eqnarray}
with the previously introduced notation.
Such a measured quantity ranges in $ [0,1] $, it is maximal when layers $ \alpha $ and $ \beta $ are identical - that is, 
if there is complete similarity between those two layers - and minimal when they are fully uncorrelated; in this perspective, it evaluates the tendency of nodes to share 
links in different layers.

However, this quantity is uninformative without a comparison with the value of binary multiplexity obtained when considering a null model. We may indeed measure
high values of multiplexity between two layers due to the possibly large observed values of density, without any significant distinction between real correlation and overlap imposed 
by the presence of many links in each layer (thus forcing an increase in the overlap itself).

Furthermore, we cannot draw a clear conclusion about the amount of correlation between layers by just looking at the observed value, since such a measure is not 
universal and, for instance, no comparison between different multiplexes can be done based on the raw multiplexity.
As we said before, such a quantity is indeed significantly dependent on the density of the layers composing the multi-network. 

We therefore introduce the following rescaled quantity~\cite{Squartini5, Garlaschelli3}:
\begin{eqnarray}
\mu^{\alpha, \beta}_{bin} = \frac{m^{\alpha, \beta}_{bin}  - \langle m^{\alpha, \beta}_{bin} \rangle}{1  - \langle m^{\alpha, \beta}_{bin} \rangle}
\label{def_mu_bin}
\end{eqnarray}
where $ m^{\alpha, \beta}_{bin} $ is the value measured for the observed real-world multiplex and $ \langle m^{\alpha, \beta}_{bin} \rangle $ is the value 
expected under the chosen null model. This rescaled quantity is now directly informative about the real correlation between layers: in this context, 
since $ \mu^{\alpha, \beta} \in [ -1, 1] $, positive values represent positive correlations, while negative values are associated to anticorrelated pairs of layers; 
furthermore, pairs of uncorrelated layers show multiplexity values comparable with 0.
Indeed, when the random graph is considered as a null model, the previous quantity (\ref{def_mu_bin}) is actually  
the correlation coefficient between the entries of the adjacency matrix referred to any two layers $ \alpha $ and $ \beta $ of a multi-level graph.

In order to compute $ \mu^{\alpha, \beta}_{bin} $ we should then calculate the expected multiplexity under the chosen null model, that is:
\begin{eqnarray}
\langle m^{\alpha, \beta}_{bin} \rangle = \frac{2 \sum_{i < j}^{} \langle \min \{ a_{ij}^{\alpha}, a_{ij}^{\beta} \} \rangle }{ \langle L_{TOT}^{\alpha} \rangle + \langle L_{TOT}^{\beta} \rangle }
\label{expected_bin_mul}
\end{eqnarray}
However, since both the considered null models preserve the average number of links in each layer as constraint, we have just to evaluate the analytical expression for the expected 
value of the minimum of two variables; for the random graph it is easy to show that (calculations are reported in the Appendix):

%%% minimo binario RG:
\begin{eqnarray}
\langle \min \{ a_{ij}^{\alpha}, a_{ij}^{\beta} \} \rangle_{RG} = p^{\alpha} p^{\beta}
\end{eqnarray}

where we define $ p^{\alpha} $ as the fraction of links actually present in that layer, as we have already done before:
\begin{eqnarray}
p^{\alpha} = \frac{L_{TOT}^{\alpha}}{N(N - 1)/2}
\end{eqnarray}

Similarly, for the binary configuration model:
%%% minimo binario CM:
\begin{eqnarray}
\langle \min \{ a_{ij}^{\alpha}, a_{ij}^{\beta} \} \rangle_{BCM} = p_{ij}^{\alpha} p_{ij}^{\beta}
\end{eqnarray}

It is now possible to compute the analytical expression for the rescaled multiplexity. We obtain for the random graph:
\begin{eqnarray}
\mu^{\alpha, \beta}_{RG} = \frac{2 \sum_{i < j}^{} \left( \min \{ a_{ij}^{\alpha}, a_{ij}^{\beta} \} - p^{\alpha} p^{\beta}\right)}{\sum_{i < j}^{} \left( a_{ij}^{\alpha} + a_{ij}^{\beta} - 2 p^{\alpha} p^{\beta} \right)}
\label{mu_bin_RG}
\end{eqnarray}
and for the binary configuration model:
\begin{eqnarray}
\mu^{\alpha, \beta}_{BCM} = \frac{2 \sum_{i < j}^{} \left( \min \{ a_{ij}^{\alpha}, a_{ij}^{\beta} \} - p_{ij}^{\alpha} p_{ij}^{\beta}\right)}{\sum_{i < j}^{} \left(a_{ij}^{\alpha} + a_{ij}^{\beta} - 2 p_{ij}^{\alpha} p_{ij}^{\beta}\right)}
\label{mu_bin_BCM}
\end{eqnarray}
%%%%%%%%%%%%%%
%%z-score

As we have already said, such rescaled quantities provide proper information about the similarity between layers of a multiplex, by evaluating the correlations measured in a 
real network with respect to what we would expect, on average, for an ensemble of multi-level networks sharing only some of the topological properties of the observed one. However,
we cannot understand, from the obtained values of multiplexity itself, whether the observed value of $ m_{BCM} $ is actually compatible with the expected one,
as $ \mu_{BCM} $ (and the correspondig value related to the random graph) does not provide any information about the standard deviation associated to the 
expected value of multiplexity.

In order to solve this issue, we introduce the z-score associated to the previously defined multiplexity:

\begin{eqnarray}
z \left[m^{\alpha, \beta}\right] = \frac{m^{\alpha, \beta}  - \langle m^{\alpha, \beta} \rangle}{\sigma \left[m^{\alpha, \beta} \right]}
\end{eqnarray}
where $ m^{\alpha, \beta} $ is the measured multiplexity between a given pair of layers on the real-world network, $ \langle m^{\alpha, \beta} \rangle $ is the value 
expected under the chosen null model and $ \sigma[m^{\alpha, \beta}] $ is the related standard deviation. The z-score, therefore, shows by how many standard deviations the observed value
of multiplexity differs with respect to the expected one for any pair of layers. In particular, in the binary case such a quantity becomes:
\begin{eqnarray}
z \left[m^{\alpha, \beta}\right] = \frac{\sum_{i < j}^{} min \{ a_{ij}^{\alpha}, a_{ij}^{\beta}\} - \sum_{i < j}^{} \langle min \{ a_{ij}^{\alpha}, a_{ij}^{\beta}\} \rangle}{\sigma \left[\sum_{i < j}^{} min \{ a_{ij}^{\alpha}, a_{ij}^{\beta}\} \right]}
\end{eqnarray}

Interestingly, not only the expected value, but even the standard deviation can be calculated analytically (calculations are reported in the Appendix). Indeed:
\begin{eqnarray}
\sigma^2 \left[\min \{ a_{ij}^{\alpha}, a_{ij}^{\beta}\} \right] = \langle {\min}^2 \{ a_{ij}^{\alpha}, a_{ij}^{\beta} \} \rangle - \langle \min \{ a_{ij}^{\alpha}, a_{ij}^{\beta} \} \rangle^2 \nonumber \\
\end{eqnarray}
and, since $ a_{ij}^{\alpha} = 0,1 $, when the configuration model is considered:
\begin{eqnarray}
\langle {\min}^2 \{ a_{ij}^{\alpha}, a_{ij}^{\beta} \} \rangle = p_{ij}^{\alpha} p_{ij}^{\beta}
\label{min_quadro}
\end{eqnarray}
Extending (\ref{min_quadro}) to the random graph is straightforward.

Hence, the z-score associated to the binary multiplexity according to the binary random graph is given by:
\begin{eqnarray}
z_{RG} = \frac{\sum_{i < j}^{} \min \{ a_{ij}^{\alpha}, a_{ij}^{\beta}\} - \sum_{i < j}^{} p^{\alpha}p^{\beta}}{\sqrt{\sum_{i < j}^{} \left[ p^{\alpha}p^{\beta} - \left(p^{\alpha}p^{\beta}\right)^2 \right]}}
\label{z_mu_bin_RG}
\end{eqnarray}
where we used the previous definitions for $ p^{\alpha} $ and $ p^{\beta} $, while for the binary configuration model:
\begin{eqnarray}
z_{BCM} = \frac{\sum_{i < j}^{} \min \{ a_{ij}^{\alpha}, a_{ij}^{\beta}\} - \sum_{i < j}^{} p_{ij}^{\alpha}p_{ij}^{\beta}}{\sqrt{\sum_{i < j}^{} \left[ p_{ij}^{\alpha}p_{ij}^{\beta} - \left(p_{ij}^{\alpha}p_{ij}^{\beta}\right)^2 \right] }}
\label{z_mu_bin_BCM}
\end{eqnarray}

We should point out that such z-scores should in principle be defined only if the associated property (in this case, $ \mu_{BCM} $) is normally distributed;
nevertheless, even if such assumption does not occur, they provide important information about the consistency between observed and randomized values.
It is worth saying that these z-scores provide a different kind of information with respect to the previous multiplexities. Mathematically, the only correlation between, for example, 
$ \mu_{BCM} $ and the corresponding $ z_{BCM} $ is the sign concordance; furthermore, the z-score is useful in order to understand whether, for 
instance, values of multiplexity close to 0 are actually comparable with 0, so that we can consider those two layers as uncorrelated, or they are instead significantly 
unexpected, although very small.
In this perspective, we should not expect a particular relation between such two variables $ \mu_{BCM} $ and $ z_{BCM} $ (or, respectively, $ \mu_{RG} $ and $ z_{RG} $).

%%%%%%%%%%%%%%%%%%%%%%%%%%%%
\subsection{Weighted multiplexity}

We now extend the previous definitions to weighted multiplex networks. We define the weighted multiplexity as:
%%%%%%%%%%%%%%%%%%%%%%%%%%%%%%%%
\begin{eqnarray}
m^{\alpha, \beta}_{w} = \frac{2 \sum_{i < j}^{} min \{ w_{ij}^{\alpha}, w_{ij}^{\beta}\}}{W_{TOT}^{\alpha} + W_{TOT}^{\beta}}
\label{m_wei}
\end{eqnarray}
where $ w_{ij}^{\alpha} $ represents the weight of the link between nodes $ i $ and $ j $ in layer $ \alpha $ and $ W_{TOT}^{\alpha} $ is the total 
weight related to the links in that layer.
Similarly to the binary case, this quantity ranges in $ [0,1] $ and it is maximal when layers $ \alpha $ and $ \beta $ are 
fully correlated. However, (\ref{m_wei}) provides a refinement of the previous binary multiplexity, since it takes into account also the weights associated to the existing links
for the evaluation of the overlap between layers of the system.

We should again compare such a measured value, however, with the value of weighted multiplexity obtained for a proper benchmark, since also in this case the distribution 
of weights across the different layers may affect the raw measure, thus making impossible a comparison between the observed weighted overlaps associated to different systems. 

Therefore, we define the following rescaled quantity:
\begin{eqnarray}
\mu^{\alpha, \beta}_{w} = \frac{m^{\alpha, \beta}_{w}  - \langle m^{\alpha, \beta}_{w} \rangle}{1 - \langle m^{\alpha, \beta}_{w} \rangle}
\label{mu_wei}
\end{eqnarray}
where $ \langle m^{\alpha, \beta}_{w} \rangle $ is the value measured for the observed real-world network and $ \langle m^{\alpha, \beta}_{w} \rangle $ is the value 
expected under the considered reference model. Again, the sign of $ \mu^{\alpha, \beta}_{w} $ is then directly informative about the weighted correlations existing between layers.

We should point out that, in this case, we cannot recover a Pearson correlation coefficient from (\ref{mu_wei}) even 
when the weighted random graph is considered as a reference; hence the previous relation just derives from a 
generalization of the methodology exploited to define the rescaled binary multiplexity, as shown in~\cite{Squartini5}.

In this context, the expected value of weighted multiplexity is given by:
\begin{eqnarray}
\langle m^{\alpha, \beta}_{w} \rangle = \frac{2 \sum_{i < j}^{} \langle \min \{ w_{ij}^{\alpha}, w_{ij}^{\beta} \} \rangle }{ \langle W_{TOT}^{\alpha} \rangle + \langle W_{TOT}^{\beta} \rangle }
\label{expected_wei_mul}
\end{eqnarray}
However, since both the weighted random graph and the weighted configuration model preserve the average total weight associated to the links in each layer as constraint, 
also in this case we just need to evaluate the analytical expression for the expected value of the minimum of two variables; 
the only difference with respect to the binary description
is related to a change in the underlying probability distribution~\cite{Squartini5}. Some simple calculations, reported in the Appendix, lead to:
%%%%%%%%%%%%%%%%%%%%%%%%%%%%%%%
%%% minimo pesato:
\begin{eqnarray}
\langle \min \{ w_{ij}^{\alpha}, w_{ij}^{\beta} \} \rangle_{WRG} = \frac{p^{\alpha} p^{\beta}}{1 - p^{\alpha} p^{\beta}}  
\end{eqnarray}
for the weighted random graph, where we define $ p^{\alpha} $, according to the likelihood maximization, as:
\begin{eqnarray}
p^{\alpha} = \frac{W_{TOT}^{\alpha}}{W_{TOT}^{\alpha} + N(N - 1)/2},
\end{eqnarray}
while for the weighted configuration model we find:
\begin{eqnarray}
\langle \min \{ w_{ij}^{\alpha}, w_{ij}^{\beta} \} \rangle_{WCM} = \frac{p_{ij}^{\alpha} p_{ij}^{\beta}}{1 - p_{ij}^{\alpha} p_{ij}^{\beta}}  
\end{eqnarray}

We can now compute the analytical expression for the rescaled multiplexity, according to both the chosen null models. We obtain for the random graph:
\begin{eqnarray}
\mu^{\alpha, \beta}_{WRG} = \frac{2 \sum_{i < j}^{} \left(\min \{ w_{ij}^{\alpha}, w_{ij}^{\beta} \} - \frac{p^{\alpha} p^{\beta}}{1 - p^{\alpha} p^{\beta}}\right)}{\sum_{i < j}^{} \left(w_{ij}^{\alpha} + w_{ij}^{\beta} - 2 \frac{p^{\alpha} p^{\beta}}{1 - p^{\alpha} p^{\beta}}\right)}
\label{mu_wei_WRG}
\end{eqnarray}
and for the binary configuration model:
\begin{eqnarray}
\mu^{\alpha, \beta}_{WCM} = \frac{2 \sum_{i < j}^{} \left(\min \{ w_{ij}^{\alpha}, w_{ij}^{\beta} \} - \frac{p_{ij}^{\alpha} p_{ij}^{\beta}}{1 - p_{ij}^{\alpha} p_{ij}^{\beta}}\right)}{\sum_{i < j}^{} \left(w_{ij}^{\alpha} + w_{ij}^{\beta} - 2 \frac{p_{ij}^{\alpha} p_{ij}^{\beta}}{1 - p_{ij}^{\alpha} p_{ij}^{\beta}}\right)}
\label{mu_wei_WCM}
\end{eqnarray}
with the previously defined notation.
%%%%%%%%%%%%%%%
%% z-score

Furthermore, we can extend to the weighted case the analysis of the z-scores associated to the values of multiplexity as defined in (\ref{mu_wei}). We then find for the 
weighted random graph (explicit calculations reported in the Appendix):
\begin{eqnarray}
z_{WRG} = \frac{\sum_{i < j}^{} \min \{ w_{ij}^{\alpha}, w_{ij}^{\beta}\} - \sum_{i < j}^{} \frac{p^{\alpha}p^{\beta}}{1 - p^{\alpha}p^{\beta}}}{\sqrt{\sum_{i < j}^{} \frac{p^{\alpha}p^{\beta}}{\left(1 - p^{\alpha}p^{\beta}\right)^2}}}
\label{z_mu_wei_WRG}
\end{eqnarray}
where we used the previous definitions for $ p^{\alpha} $ and $ p^{\beta} $, while for the binary configuration model:
\begin{eqnarray}
z_{WCM} = \frac{\sum_{i < j}^{} \min \{ w_{ij}^{\alpha}, w_{ij}^{\beta}\} - \sum_{i < j}^{} \frac{p_{ij}^{\alpha}p_{ij}^{\beta}}{1 - p_{ij}^{\alpha}p_{ij}^{\beta}}}{\sqrt{\sum_{i < j}^{} \frac{p_{ij}^{\alpha}p_{ij}^{\beta}}{\left(1 - p_{ij}^{\alpha}p_{ij}^{\beta}\right)^2}}}
\label{z_mu_wei_WCM}
\end{eqnarray}

\section{Results}

\subsection{Data}

We validate our definitions applying them to two different real-world multiplexes: the International Trade Network and the European Airport Network.
%%%%% ITN
In particular, we analyze the International Trade Network, also known as World Trade Web, as provided by the BACI database~\cite{Baci}. 
The data provide information about import and export between 207 countries in 2011 and 
turns out to have a straightforward representation in terms of multi-layered network~\cite{Barigozzi}; it is indeed possible to disaggregate the global 
trade between any two countries into
the import and export in a given commodity, so that the global trade system can be thought as the superposition of all the layers.
The network is then composed by 207 countries and 96 different commodities, according to the standard international classification HS1996~\cite{Comtrade} (the list of 
commodities is reported in the Appendix).
While the aggregated network shows
a density equal to about $ 63 \% $, the various layers are characterized by densities from $ 6 \% $ (related to trade in silk) to $ 45 \% $ (for import-export of mechanical appliances
and parts thereof). Such heterogeneity may suggest that a multiplex analysis is therefore required.
Interestingly, in this case each of the layers is represented by a weighted network, where the weight associated to any link in a layer stands for the amount of money exchanged 
by a given pair of countries in that layer (i.e., commodity).

%%%%% airport
The second multi-level network we analyze is the European Airport System. Here, the european airports represent the nodes, and the layer stand for the different airline companies 
active in Europe; hence, in a given layer a link between two nodes is present if there exists at least one direct flight between those two airports, operated by that airline. 
The dataset we consider has been provided by OpenFlight~\cite{Openflight}, a free on-line platform  supplying information about the flights taking place all over the world.
As we said before, we focus on the european network: 669 airports are thus considered, reached by flights operated by 171 companies (listed in the Appendix). 
Unlike the International Trade Network,
such a system can only be studied from an unweighted perspective, according to the chosen dataset. Moreover, both the aggregated network and the various layers show densities which are
significantly lower than those observed in the previous dataset: indeed, in this system the aggregated network exhibits a density of about $ 2 \% $, since a given airport has 
a number of connections with other cities which is very limited with respect to the global number of european airports.

%% dire che gli hubs sono tutti diversi e che pochi nodi sono condivisi da molti layers
We must therefore point out once more that the introduction of proper null models and the use of rescaled quantities allow us to appropriately compare the results obtained for
such different networks, independently from the size and the density of the considered systems.

\subsection{Binary analysis}
%%%%%%%%%%%%%%%%%%%%%%%%%

\begin{figure*}[htbp]  %%% con asterisco per avere su due colonne
\begin{center}
\includegraphics[width=1.0\textwidth]{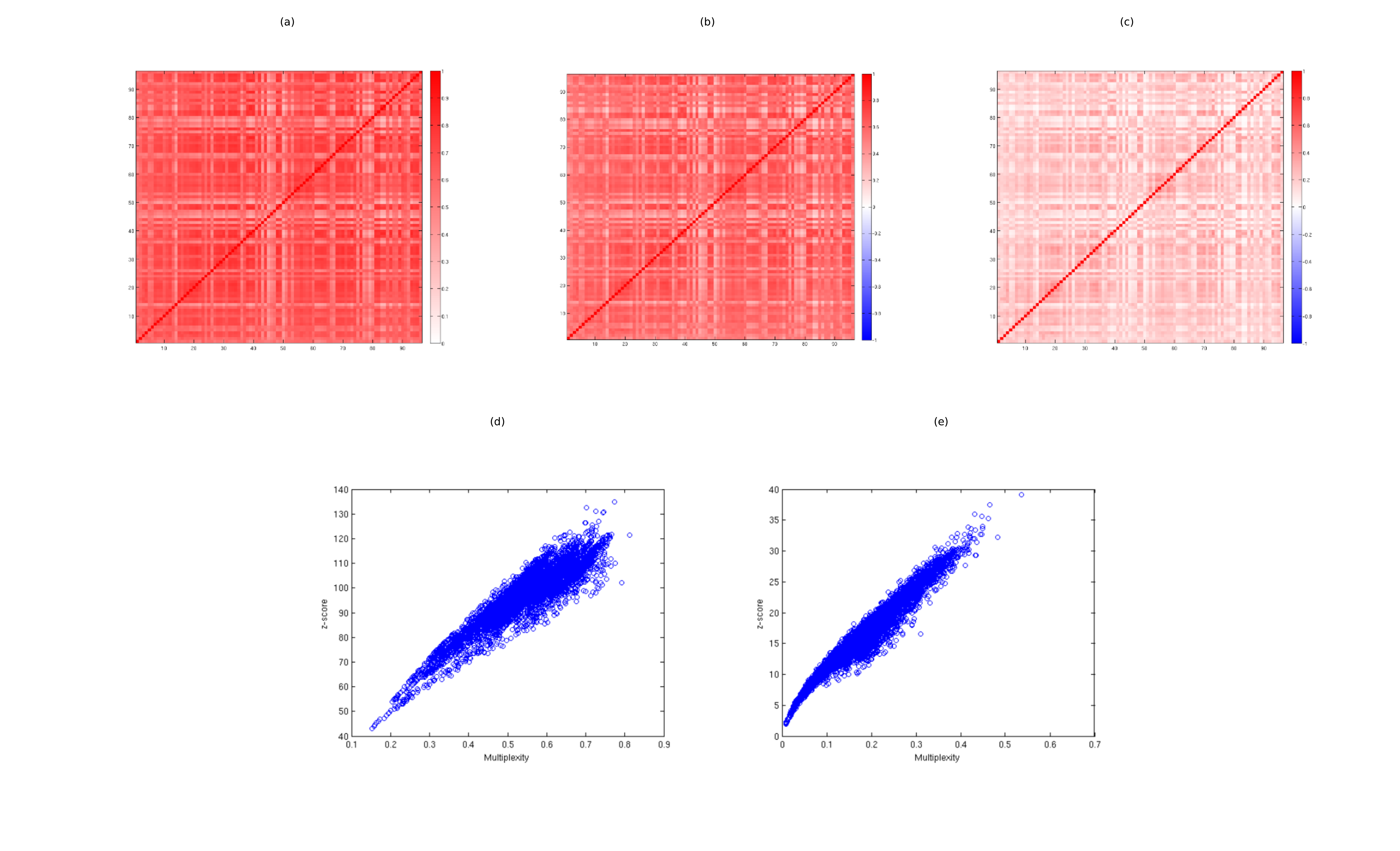}
\end{center}
\caption{Analysis of the binary correlations between layers of the International Trade Network in 2011. Top panels: multiplexity color-coded matrices; entries represent values of
$ m_{bin} $ (a), $ \mu_{RG} $ (b) and $ \mu_{BCM} $ (c) for any pair of layers (commodities).
Bottom panels: scatter plots of binary multiplexity values vs the corresponding z-score for each pair of layers, respectively for random graph (d) and binary configuration 
model (e).}
\label{fig:subplot_bin_BACI}
\end{figure*}

The implementation of the concept of multiplexity to different networks can lead to completely divergent results, according to the structural features of the considered 
systems. Indeed, the application of (\ref{m_bin}) to the International Trade Network leads to the color-coded multiplexity matrix shown in 
Figure \ref{fig:subplot_bin_BACI}(a). Such an array generally shows very high overlaps between layers, i.e. between different classes of commodities, pointing out that
usually each country tends to import from or export to the same set of countries almost independently from the traded items; this is true in particular for most of the 
edible products (layers characterized by commodity codes ranging from 1 to 22, as listed in the Appendix).
In order to have a complete picture of the analyis of correlations between layers of the considered systems, we have to compare our findings with the overlaps
expected for multiplexes having only some of the properties in common with the observed ones.
The simplest benchmark, as well as the most widely used, is the random graph, which discards, as we said, any kind of heterogeneity in the degree distributions of 
the layers. When we apply (\ref{mu_bin_RG}) to the International Trade Network, we obtain the multiplexity matrix shown in Figure \ref{fig:subplot_bin_BACI}(b). It clearly
shows that most of the correlations are still present: this layer-homogeneous null model, together with the presence of comparable densities across the various
layers, does not significantly affect the expected overlaps.
So far, we have discarded heterogeneity in our null models. However, this can considerably affect the significance of our findings.
Therefore, we introduce heterogeneity in the degree distribution within the reference model by means of the previously defined configuration model.
This way, it is actually possible to detect only the non-trivial correlations, therefore discarding all the overlaps simply due to the possibly high density of the layers, 
that would otherwise increase the observed interrelations even if no real correlation is actually present.

\begin{figure*}[htbp]  %%% con asterisco per avere su due colonne
\begin{center}
\includegraphics[width=1.0\textwidth]{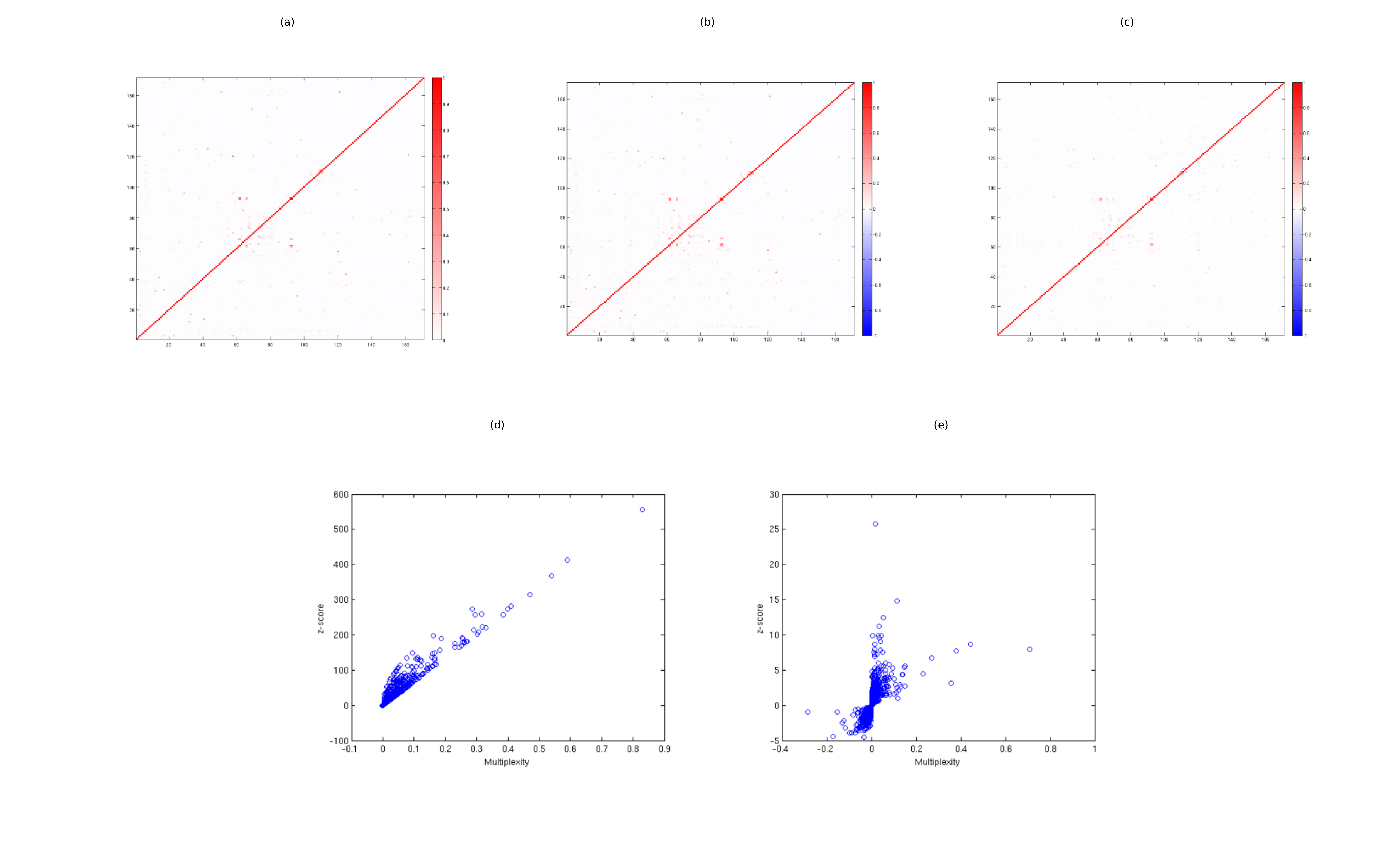}
\end{center}
\caption{Analysis of the binary correlations between layers of the European Airport Network. Top panels: multiplexity color-coded matrices; entries represent values of
$ m_{bin} $ (a), $ \mu_{RG} $ (b) and $ \mu_{BCM} $ (c) for any pair of layers (airlines).
Bottom panels: scatter plots of binary multiplexity values vs the corresponding z-score for each pair of layers, respectively for random graph (d) and binary configuration 
model (e).}
\label{fig:subplot_bin_flights}
\end{figure*}

This is exactly what happens when the World Trade Network is analyzed. Indeed, as shown in Figure \ref{fig:subplot_bin_BACI}(c), we find out that a 
significant amount of the binary overlap observed in this network is actually due to the information included in the degree sequence of the various layers, 
rather than to a real correlation between layers. This method is therefore able to detect the really meaningful similarity between layers, discarding the 
trivial overlap caused by the presence, for instance, of nodes having a high number of connections in most of the layers. 
This non-significant overlap is thus destroyed by our procedure.  
Such observations clearly show that the random graph is therefore not the most proper reference model in order to obtain an appropriate representation 
of crucial properties of such multi-level systems. 
\begin{figure*}[htbp]  %%% con asterisco per avere su due colonne
\centering
%\begin{center}
\includegraphics[width=17cm,height=19cm]{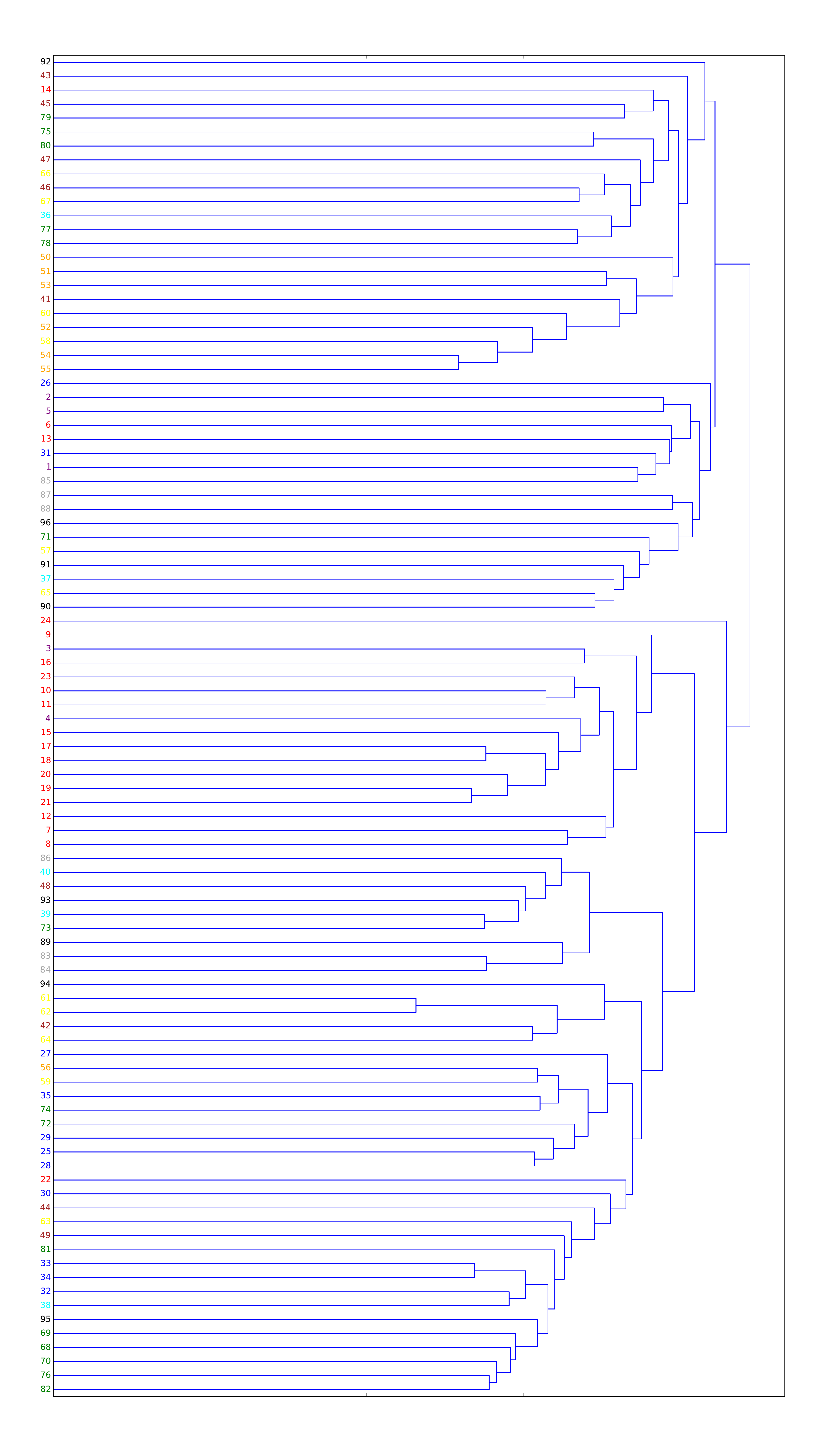}
%\includegraphics[width=1.0\textwidth]{dendro_BACI2011_bin_undir_resc_rotated.pdf}
%\end{center}
\caption{Dendrogram of commodities traded in 2011 as obtained applying the Average Linkage Clustering Algorithm to the binary rescaled multiplexity $ \mu_{BCM} $; colors of the leaves 
represent different classes of commodities, as reported in the Appendix.}
\label{fig:dendro_BACI}
\end{figure*} 
In order to have a better understanding of the correlations between layers, it is possible to implement a hierarchical clustering procedure starting from each of the 
multiplexity matrices in Figure \ref{fig:subplot_bin_BACI}~\cite{Mantegna}. However, we have to define a notion of distance between layers, starting from our notion of 
correlation. We can define a distance $ d^{\alpha, \beta} $ between any pair of commodities in the following way:
\begin{eqnarray}
d^{\alpha, \beta} = \sqrt{ \frac{1 - \mu_{BCM}^{\alpha, \beta}}{2}}.
\end{eqnarray}
Hence, the maximum possible distance $ d^{\alpha, \beta} $ between any two layers is 1 (when layers $ \alpha $ and $ \beta $ show multiplexity $ \mu_{BCM}^{\alpha, \beta} = 1 $), 
while the minimum one is 0 (corresponding to $ \mu_{BCM}^{\alpha, \beta} = 1 $).
We can therefore represent the layers of the multiplex as the leaves of a taxonomic tree, where highly correlated communities meet at a branching point which is closer to 
baseline level. In Figure \ref{fig:dendro_BACI} we show the dendrogram obtained by applying the Average Linkage Clustering Algorithm to the matrix, shown in Figure 
\ref{fig:subplot_bin_BACI}(c), representing values of multiplexity $ \mu_{BCM} $. We can see that some groups of similar commodities are clearly visible (for instance, the 
group of edible commodities is easily identified), while in other cases apparently distant commodities are grouped together, pointing out that some unexpected 
correlations are present.
Similar dendrograms can be designed starting from the matrices in Figures \ref{fig:subplot_bin_flights}(a) and \ref{fig:subplot_bin_flights}(b). 

A completely different behaviour can be observed when, instead, the European Airport System is considered.
Indeed, low values of multiplexity observed for such a network (Figure \ref{fig:subplot_bin_flights}(a)) illustrate nearly no overlap between most of the layers: 
this highlights the well-known tendency of airline companies to avoid superpositions between routes with other airlines.
In Figure \ref{fig:subplot_bin_flights}(b) we show the residual correlations obtained after the application
of the random graph: almost no difference can be perceived with respect to Figure \ref{fig:subplot_bin_flights}(a), since the expected overlap in this case is very small, due
to the very low densities of the various layers. We should point out that the random graph is not a 
proper reference model for this real-world network, since the assumption of uniformity in the degrees of the different nodes (i.e., airports) is actually far from the observed 
structure of such a system, as we will highlight later.
Nevertheless, in Figure \ref{fig:subplot_bin_flights}(c) we show that, at first glance, the adoption of the configuration model looks not strictly required 
when the European Airport Network is considered, except for a more suitable mathematical approach, since the overall matrix looks apparently similar to the 
previous Figure \ref{fig:subplot_bin_flights}(b). The presence of a larger number of negative values of 
multiplexity, however, highlights once more the anti-correlated character of such a system, and this crucial structural property of the airport multiplex network was not 
clearly revealed by the application of the random graph.

In this case, a dendrogram designed form matrices reported in Figure \ref{fig:subplot_bin_flights} would not be meaningful, since most of the layers meet at a single root level, 
due to the very low correlation observed between them. 

\subsubsection{Binary z-scores}

%% ITN
As we said in the previous Section, color-coded multiplexity matrices are useful in order to detect the meaninful correlations between layers in a multiplex, but they do not 
supply any information about the discrepancy of the observed values from the corresponding expected ones. Hence, the introduction of suitable z-scores associated to the 
previously defined quantities is required. Moreover, it is worth reminding that the information provided by (\ref{mu_bin_BCM}) (respectively (\ref{mu_bin_RG}) for the random graph) 
is not necessarily connected to that supplied by (\ref{z_mu_bin_BCM}) (respectively, (\ref{z_mu_bin_RG}))
Indeed, while the multiplexity by itself detects the degree of correlation between layers of a multi-level network, the corresponding z-scores reveal how significant 
those values actually are with respect to our expectations.

In Figure \ref{fig:subplot_bin_BACI}(d) we show, for the International Trade Network, the scatter plot of the values of binary multiplexity vs the corresponding z-scores, 
after comparing the observed 
values with the expected ones under random graph. We show that observed very large values of z-scores reveal a high significance of the previously obtained overlaps; such a 
consideration therefore points out that even the pairs of layers showing low (but positive) values of multiplexity cannot actually be considered as uncorrelated. Furthermore, 
a clear correlation between $ \mu_{RG} $ and $ z_{RG} $ can be observed, thus large values of binary multiplexity correspond to large z-scores, and vice-versa.

Similar considerations can be done when the binary configuration model is considered as a beachmark.
Indeed, as we show in Figure \ref{fig:subplot_bin_BACI}(e), a large correlation between $ \mu_{BCM} $ and $ z_{BCM} $ is still present when we consider the International
Trade Network; moreover, since almost all the z-scores are higher than the widely used critical value $ z^{*}_{BCM} = 2 $ (so that almost no pair of layers
shows a multiplexity lying within 2 standard deviations form the expected value), we highlight that most of the pairs therefore exhibit unexpectedly high 
correlations with respect to the corresponding average value obtained when randomizing the real-world layers according to the configuration model, similarly to what 
we found before for the random graph.

However, if we look at the absolute values of such z-scores, we observe that the significance of the values of multiplexity under random graph ($ \mu_{RG} $) is generally 
much higher than that measured under configuration model ($ \mu_{BCM} $). 

%% EAN
A different trend can be observed when the European Airport Network is taken into account (Figure \ref{fig:subplot_bin_flights}(d)). 
Indeed, it is still clear a high correlation between values of multiplexity and their respective z-scores when the random graph is considered.
However, many z-scores associated to
multiplexities close to 0, in this case, are now close to 0 themselves, therefore suggesting that many pairs of layers (i.e. airline companies) may actually
be anti-correlated rather simply uncorrelated. In this case, the adoption of a more refined null model is then crucial in order to deeply understand the structural 
properties of such a system.

When the binary configuration model is considered as benchmark, however, the analysis of the corresponding scatter plots dramatically changes.
However, as we said, these results are strongly dependent on the considered network. Indeed, Figure \ref{fig:subplot_bin_flights}(e) exhibits a completely different 
trend with respect, for instance, to the corresponding Figure \ref{fig:subplot_bin_BACI}(e) (related to the World Trade Network):
no correlation between $ \mu_{BCM} $ and $ z_{BCM} $ can be observed in this case, so that
the same value of multiplexity can be either associated to a low z-score (thus being compatible with the expected value under the chosen configuration model) or to very high z-scores 
(hence unexpectedly different from the model's expectation). 
Moreover, Figure \ref{fig:subplot_bin_flights}(e) clearly shows the sign-concordance existing between the multiplexity and the associated z-score 
that we pointed out in the Methods section. However, no other clear
trend can be inferred from such a plot,
therefore pointing out the importance of taking into account both the quantities ($ \mu_{BCM} $ and $ z_{BCM} $) in order to have a complete understanding
of the correlations between layers of a multiplex.

Furthermore, we should highlight once more that, in terms of absolute z-scores values, the significance of the values of multiplexity under random graph ($ \mu_{RG} $) is usually 
much higher than that observed after the comparison with the configuration model ($ \mu_{BCM} $), as we have already found before for the International Trade Network.

\subsection{Weighted analysis}

Since the International Trade Network is represented by a weighted multiplex, the analysis of weighted overlaps between layers of that system can be performed, in order to 
obtain more refined information about the correlations between different classes of commodities.
We should indeed point out that, for the World Trade Web, while the binary overlaps provided by (\ref{m_bin}) only supply information about the correlations between the topologies
of the various layers representing trade in different commodities, the weighted multiplexity defined in (\ref{m_wei}) is able to detect patterns of correlation
between quantities of imported and exported classes of items. In this perspective, observing high correlations is therefore more unlikely.     
This is due, mathematically, to the functional form of the definition of the multiplexity given in (\ref{m_wei}), which is significantly dependent on
the balance between weights of the corresponding links in different layers; such a property, therefore, tends to assign higher correlations to pairs of commodities 
characterized by similar global amount of trade, as we want.
\begin{figure*}[htbp]  %%% con asterisco per avere su due colonne
\begin{center}
\includegraphics[width=1.0\textwidth]{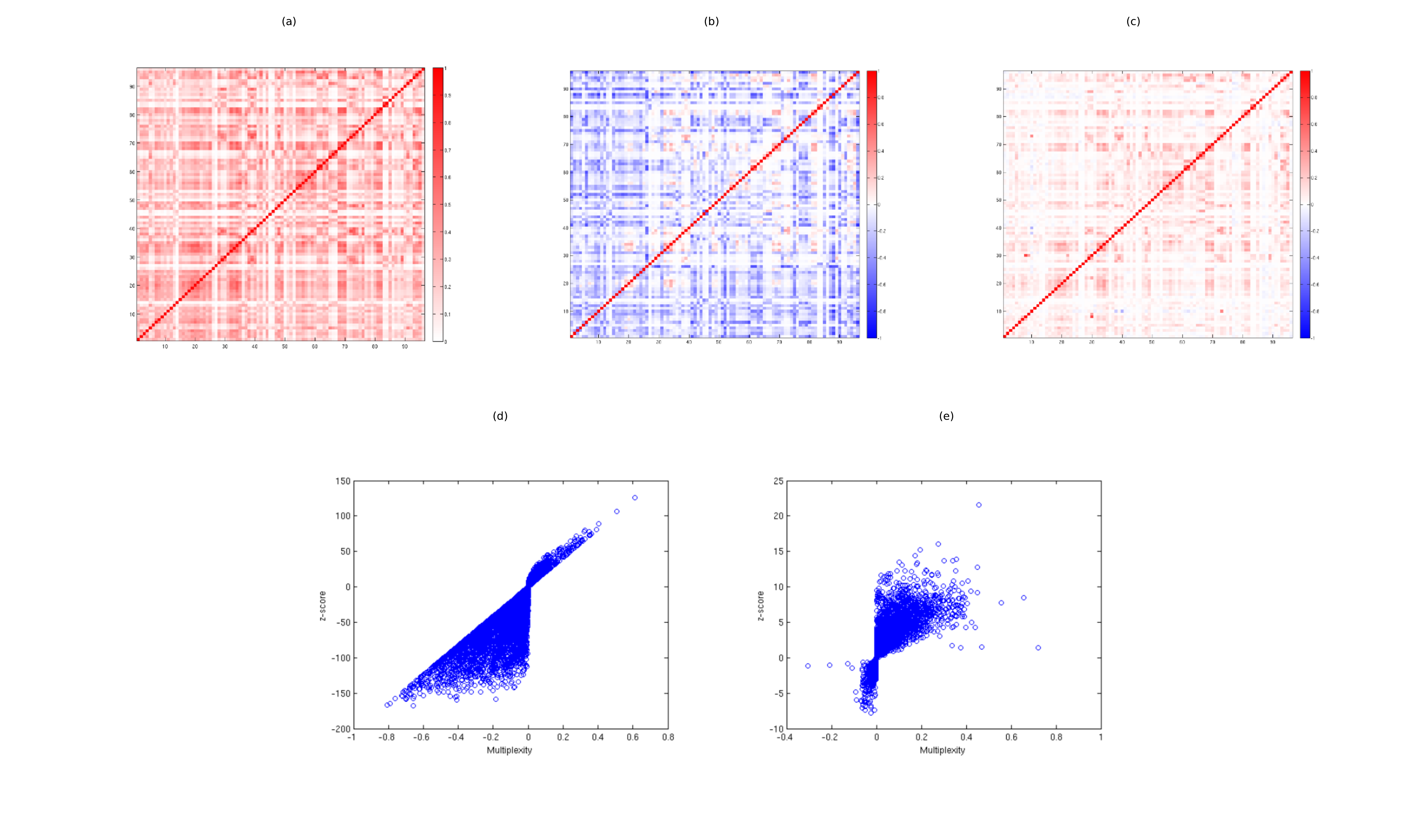}
\end{center}
\caption{Analysis of the weighted correlations between layers of the International Trade Network in 2011. Top panels: multiplexity color-coded matrices; entries represent values of
$ m_{w} $ (a), $ \mu_{WRG} $ (b) and $ \mu_{WCM} $ (c) for any pair of layers (commodities).
Bottom panels: scatter plots of weighted multiplexity values vs the corresponding z-score for each pair of layers, respectively for weighted random graph (d) and weighted configuration 
model (e).}
\label{fig:subplot_wei_BACI}
\end{figure*}
In Figure \ref{fig:subplot_wei_BACI}(a) we show the color-coded matrix associated to the raw values of weighted multiplexity as observed in the International Trade Network: clear 
correlations between different layers are still present, but a comparison with its corresponding binary matrix (shown in Figure \ref{fig:subplot_bin_BACI}(a)) explicitly reveals that, 
while some pairs of layers are significantly overlapping, several pairs of commodities are now actually uncorrelated, as expected when the weights of the links are taken into account.

In order to provide information about the relation between the observed correlations and the expected ones under a given benchmark, as a first estimate,
we apply (\ref{mu_wei_WRG}), therefore considering the corresponding weighted random graph as a reference for our real-world network.
Our findings show, in Figure \ref{fig:subplot_wei_BACI}(b), a strongly uncorrelated behavior associated to most of the pairs of commodities, in contrast with our intuitive 
expectations based on the results obtained in the binary case.

We then compare the observed multiplexity with its expected values under the weighted configuration model. Results, shown in Figure \ref{fig:subplot_wei_BACI}(c), exhibit
a completely different behavior with respect to Figure \ref{fig:subplot_wei_BACI}(b), thus highlighting once more the importance of taking into account the heterogeneity in the 
weight and degree distributions within the considered null model.
Indeed, we observe that, exploiting this more suitable reference, several pairs are still correlated, even in the weighted case, some of them are actually uncorrelated, as 
expected by looking at the corresponding binary matrix (Figure \ref{fig:subplot_bin_BACI}(c)), and only a few, with respect to the weighted random graph case, remain anti-correlated.
In general, however, the correlations in the weighted case are less noticeable, as we can see from a comparison between the matrices shown 
in Figures \ref{fig:subplot_bin_BACI}(c) and \ref{fig:subplot_wei_BACI}(c).

\subsubsection{Weighted z-scores}

We now analyze the patterns of correlations resulting from the z-scores associated to the weighted multiplexity, as defined in (\ref{z_mu_wei_WRG}) and (\ref{z_mu_wei_WCM}). 
In Figure \ref{fig:subplot_wei_BACI}(d) we show the relation between the values of weighted multiplexity for any pair of layers and the related z-score, computed with respected to the 
expected multiplexity according to the weighted random graph. The sign concordance is still clear, but the correlation between $ \mu_{WRG} $ and $ z_{WRG} $ is much less
sharp with respect to the corresponding binary case, especially for negative values of multiplexity. 

Even more so, such a weak correlation between weighted multiplexity and the corresponding z-score completely disappears when the considered benchmark is the weighted 
configuration model (Figure \ref{fig:subplot_wei_BACI}(e)): in this case the same value of $ \mu_{WCM} $ may correspond to z-scores even characterized by different 
orders of magnitude, thus pointing out once more the importance of the introduction of a notion of standard deviation referred to the average $ \langle \mu_{WCM} \rangle $.
Indeed, the same value of observed multiplexity can actually be either extremely 
unexpected or in full agreement with the null model's prediction.

\subsection{Hubs distribution}

\begin{figure*}[htbp]  %%% con asterisco per avere su due colonne
\begin{center}
\includegraphics[width=1.0\textwidth]{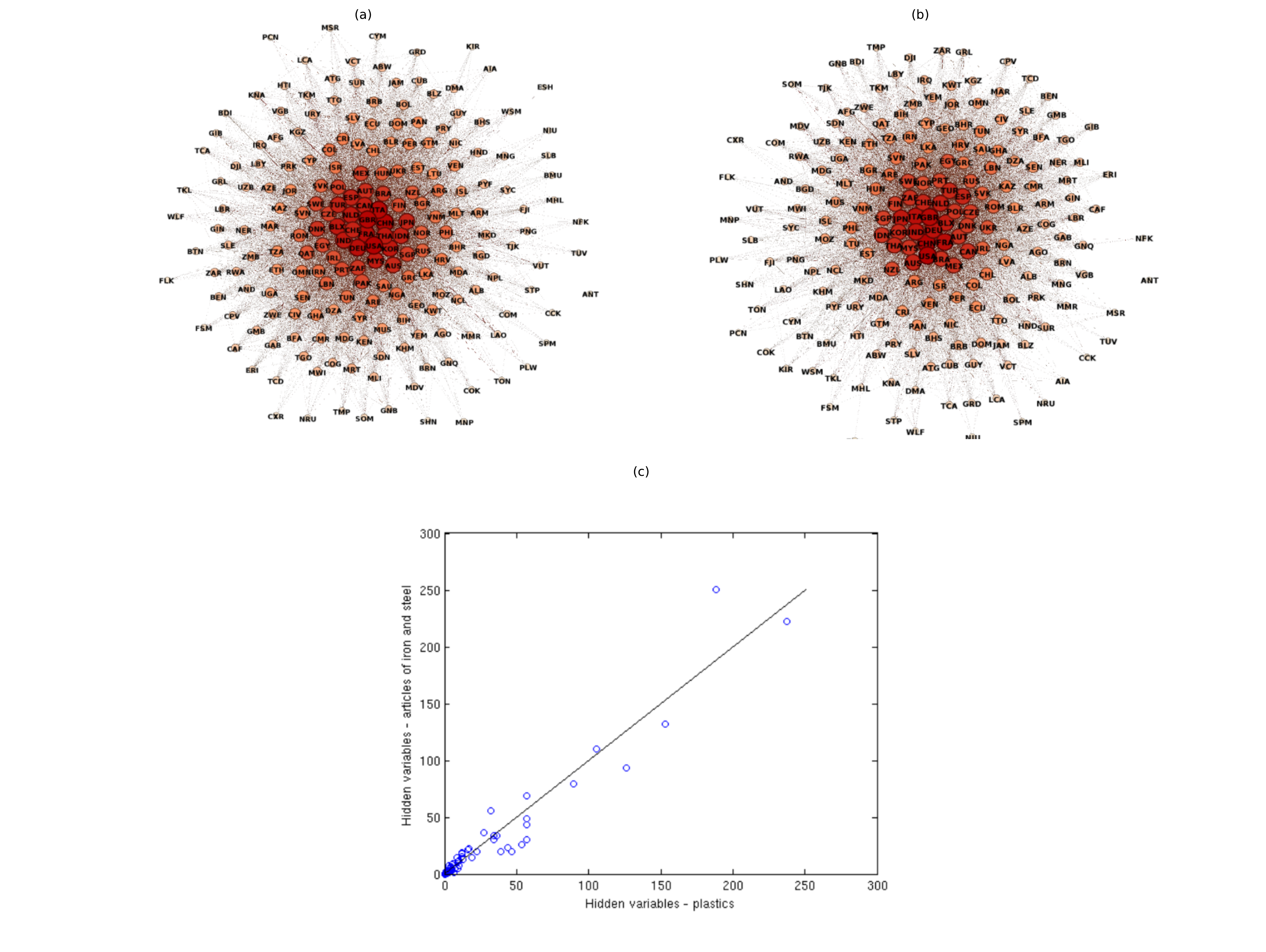}
\end{center}
\caption{Hubs distribution in the International Trade multiplex. Top panels: graphs representing two layers of the system, respectively those associated to trade in plastics (a) and 
articles of iron and steel (b); nodes represent trading countries; size and color of a node are proportional to its degree in that layer.
Bottom panel: scatter plot of the hidden variables $ x_i $ relative to each of the nodes for the same two layers; the black line represents the identity line.}
\label{fig:subplot_gephi_BACI}
\end{figure*}

Such different behaviours observed for the two considered multiplexes can be, at least partly, explained in terms of distribution of the hubs across layers. Indeed, as we show 
in Figure \ref{fig:subplot_gephi_BACI}(a) and \ref{fig:subplot_gephi_BACI}(b), generally any two layers of the World Trade Network exhibit the same set of hubs (which in this 
particular case are represented by the richest and most industrialized countries). This property, therefore, produces a higher correlation between layers, since the overlap 
is increased by the multiple presence of links in the various layers connecting nodes to the hubs.
It is possible to show that this hubs distribution, leading to the higher overlap between layers, is strongly correlated to the relation 
existing between the hidden variables $ x_i $ associated to each node in the different layers. Indeed, as shown in Figure \ref{fig:subplot_gephi_BACI}(c), for the considered pair 
of layers (but several pairs actually exhibit the same behaviour) such a trend can be clearly represented by a straight line, thus pointing out that nodes with higher $ x_i $ in 
one layer (hence, with higher probability of establishing a link with any other node in that layer) generally also have higher $ x_i $ in a different layer.

\begin{figure*}[htbp]  %%% con asterisco per avere su due colonne
\begin{center}
\includegraphics[width=1.0\textwidth]{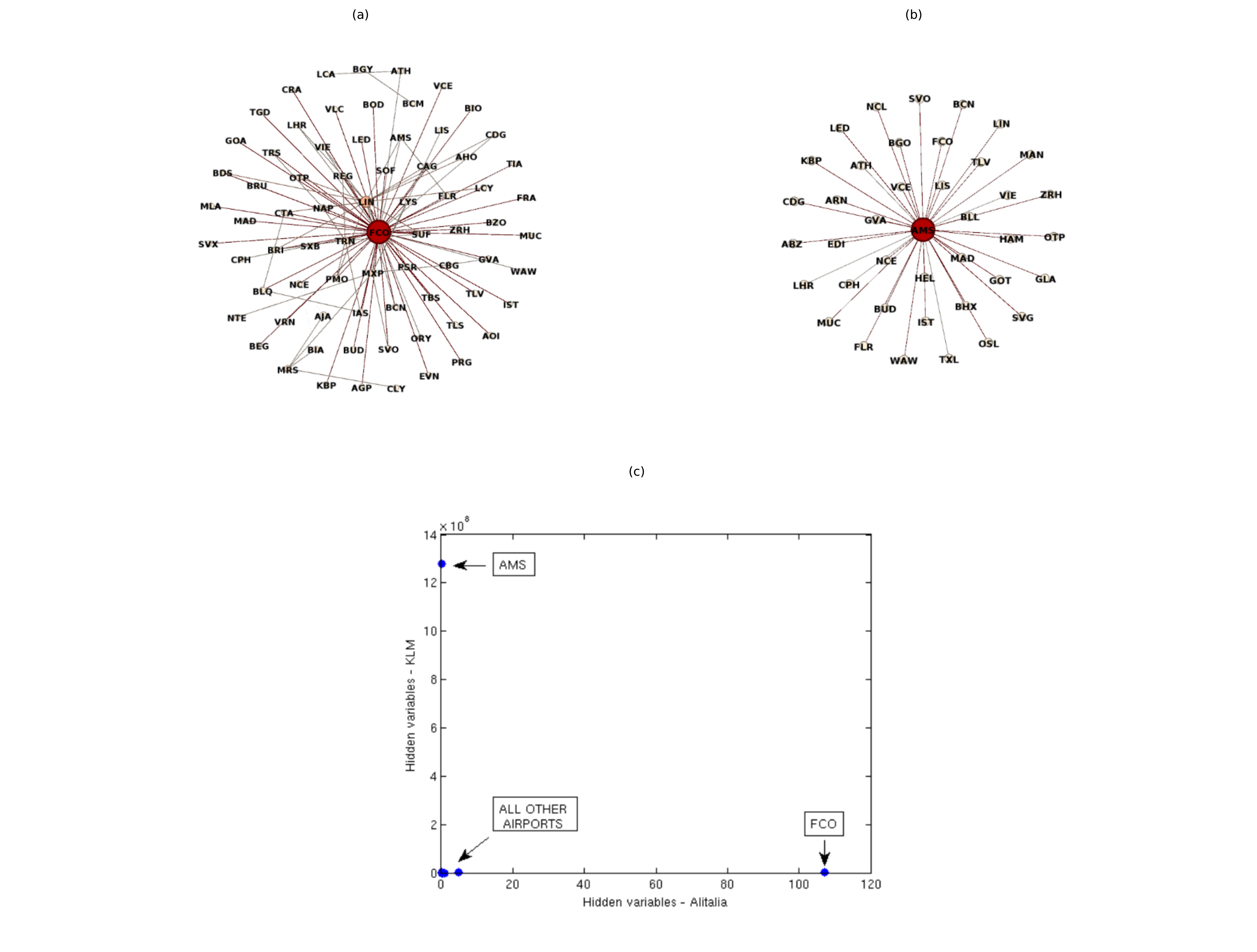}
\end{center}
\caption{Hubs distribution in the European Airport multiplex. Top panels: graphs representing two layers of the system, respectively those associated to Alitalia airline (a) and 
KLM airline (b); nodes represent european airports; size and color of a node are proportional to its degree in that layer.
Bottom panel: scatter plot of the hidden variables $ x_i $ relative to each of the nodes for the same two layers.}
\label{fig:subplot_gephi_flights}
\end{figure*}

However, when the European Airport Network is considered, an opposite trend can be observed, thus a clear explanation of the small measured overlap arises; indeed, Figures 
\ref{fig:subplot_gephi_flights}(a) and \ref{fig:subplot_gephi_flights}(b) show that in this case the layers can be approximated to star-like graphs, with a single, largely connected 
hub and several other poorly connected nodes. Though, the hub is in general different for almost any considered layer, since each airline company is based on a different airport: in the 
considered pair of layers, hubs are represented by Rome - Fiumicino airport (FCO) for Alitalia and Amsterdam - Schiphol airport (AMS) for KLM. Such a property, therefore, decreases 
significantly the overlap between layers, thus leading to the matrices previously shown in Figure \ref{fig:subplot_bin_flights}.
Similar considerations can be done when looking at Figure \ref{fig:subplot_bin_flights}(c), where the scatter plot of the hidden variables associated to the nodes in two different layers is shown. 
We observe that no linear trend can be inferred, since only the two hubs stand out from the bunch of the other airports (which are actually characterized by different 
values of $ x_i $, even though this cannot be fully appreciated). It is anyway clear that the hub of 
one layer, characterized by the highest value $ x_i $ (hence, with the highest probability of establishing a link with any other node in that layer) is a poorly connected node 
in a different layer, being characterized by a small value of $ x_i $.

In this perspective, new mechanisms explaining the growth and organization of multiplexes can be designed; in particular, proper fitness models~\cite{Garlaschelli2, Caldarelli} 
for multi-level systems may be 
introduced. According to the previous findings, the fitness parameter could be chosen as independent from the layer - hence depending only on the identity of
each vertex - if strongly 
correlated multiplexes are 
taken into account, or it may also depend on the considered layer if we are dealing with uncorrelated or anticorrelated networks

\section{Conclusions}

In the last few years the multiplex approach has revealed itself as a useful framework in order to study several real-world systems characterized by 
elementary units linked by different kinds of connection.
In this context, we have introduced new measures aiming at analyzing correlations between layers of the network, both for binary and weighted multi-graphs.
We showed that such a multiplexity is able to extract crucial information from both sparse and dense networks 
by testing it on different real-world multi-layer systems. We clearly found that a classification can be done based on the degree of
overlap between links in different layers. 
For instance, we showed that some multiplexes exhibit small overlap between links in different layers, since just a limited number of nodes are active 
in many layers, while most of them participate to one or few layers.
However, for other systems, such as the International Trade Network, most of the pairs of nodes are connected in several layers, so that such multiplexes
exhibit large overlap between layers.
Furthermore, we found that the multiplexity can also provide interesting information about the distribution of hubs across the various layers; indeed, systems 
characterized by nodes having many connections in most of the layers, such as the International Trade Network, tend to show higher values of raw binary multiplexity. On 
the other hand, in different networks, exhibiting values of multiplexity for most of the pairs of layers closed to 0, a node with a low degree in a given layer may 
represent a hub in a different layer: the European Airport Network is a clear prototype of such systems.

Our findings suggest that adopting proper null models for multi-level networks, enforcing constraints taking into account correlations between layers, is then required in order
to suitably model such real-world systems.

Further research in this direction will probably provide a better understanding of the role of local constraints in real-world multi-level systems.

\section*{Acknowledgments}

This work was supported by the EU project MULTIPLEX (contract 317532).
DG also acknowledges support from the Netherlands Organization for
Scientific Research (NWO/OCW) and the Dutch Econophysics Foundation
(Stichting Econophysics, Leiden, the Netherlands) with funds from
beneficiaries of Duyfken Trading Knowledge BV, Amsterdam, the
Netherlands.

\appendix

\section{Maximum Likelihood Method}

In order to design a suitable null model for a graph with enforced local constraints, we exploit the Maximum Likelihood Method~\cite{Squartini1}. In the binary case, when the observed 
degree sequence represents the property that we want to preserve (i.e., in the so-called configuration model), the method reduces to finding the solution to following set of $ N $ 
coupled nonlinear equation, independently for each layer $ \alpha = 1,2, \ldots, M $:
%for any node $ i $ ( $ i = 1,2,\ldots, N $) in any layer $ \alpha $:
\begin{eqnarray}
\sum_{i < j}^{} \frac{x_i^{\alpha} x_j^{\alpha}}{1 + x_i^{\alpha} x_j^{\alpha}} = k_{i}^{\alpha} \; \; \; \forall i = 1,2,\ldots, N
\end{eqnarray}
where $ k_{i}^{\alpha} $ is the observed degree of node $ i $ in layer $ \alpha $ and the unknown variables of the equation are the so-called $ N $ hidden variables 
associated to that layer.

Thus, the expected link probability $ p_{ij}^{\alpha} $ is given by, for any pair of nodes $ (i,j) $ in any layer $ \alpha $:
\begin{eqnarray}
p_{ij}^{\alpha} = \frac{x_i^{\alpha} x_j^{\alpha}}{1 + x_i^{\alpha} x_j^{\alpha}}
\end{eqnarray}
which is therefore the generalization of the equation (\ref{p_alfa_bin}) in the main text.
We can therefore see that such hidden variables $ x_i^{\alpha} $ are proportional to the expected link probability $ p_{ij}^{\alpha} $ in a given layer $ \alpha $: a higher value of 
$ x_i^{\alpha} $ will correspond to a higher expected probability of observing a link between $ i $ and any other node $ j \neq i $, and vice-versa.

Similarly, for weighted multiplexes, we can enforce the strength sequence observed in a real network on a network ensemble, thus designing a proper null model where the strength 
sequence of the considered real-world network is preserved, while the other properties are randomized. In this context, the Maximum Likelihood Method for weighted 
graphs reduces to solving a set of $ N $ coupled nonlinear equations.
For any node $ i $ in any layer $ \alpha $, we have:
\begin{eqnarray}
\sum_{i < j}^{} \frac{x_i^{\alpha} x_j^{\alpha}}{1 - x_i^{\alpha} x_j^{\alpha}} = s_{i}^{\alpha} 
\end{eqnarray}
where $ s_{i}^{\alpha} $ is the observed strength of node $ i $ in layer $ \alpha $ and the unknown variables of the equation are, again, the $ N $ hidden variables 
associated to the considered layer.

Thus, the expected link weight $ w_{ij}^{\alpha} $ is given by, for any pair of nodes $ (i,j) $:
\begin{eqnarray}
w_{ij}^{\alpha} = \frac{x_i^{\alpha} x_j^{\alpha}}{1 - x_i^{\alpha} x_j^{\alpha}}
\end{eqnarray}
hence generalizing the corresponding equation (\ref{p_alfa_wei}) of the main text. In this case, the computed hidden variables $ x_i^{\alpha} $ are proportional 
to the expected link weight $ w_{ij}^{\alpha} $ in a given layer $ \alpha $; a higher value of 
$ x_i^{\alpha} $ will therefore correspond to a higher expected link weight between $ i $ and any other node $ j \neq i $, and vice-versa

\section{Relationship with the correlation coefficient}

A possible definition \cite{Barigozzi} of correlation between layers of a multiplex builds on the standard correlation coefficient: 
\begin{eqnarray}
Corr \{ a_{ij}^{\alpha}, a_{ij}^{\beta} \} = \frac{\langle a_{ij}^{\alpha} a_{ij}^{\beta} \rangle - \langle a_{ij}^{\alpha} \rangle \langle a_{ij}^{\beta} \rangle}{\sigma_{\alpha} \sigma_{\beta}}
\label{def_corr_coeff}
\end{eqnarray}
Hence, a value of correlation
equal to $ 0 $ represents a pair of uncorrelated layers only if the probability distributions of $ a_{ij}^{\alpha} $ and $ a_{ij}^{\beta} $ are independent from the chosen node, 
that is, if all the edges in a certain layer are statistically equivalent. However, this leads to a probability of establishing a given link which is common to each pair of nodes, and this is the 
assumption behind the random graph. 

In this context, it is then possible to show that, when the binary random graph is taken into consideration, our novel measure of multiplexity can be reduced to the usual definition 
of correlation coefficient. Indeed, we have:
\begin{eqnarray}
\langle a_{ij}^{\alpha} a_{ij}^{\beta} \rangle & = & \frac{ 2 \sum_{i < j}^{} a_{ij}^{\alpha} a_{ij}^{\beta} }{N (N - 1)} = \nonumber \\
                                               & = &  \frac{ 2 \sum_{i < j}^{} \min \{ a_{ij}^{\alpha}, a_{ij}^{\beta} \} }{L_{\alpha} + L_{\beta}} \frac{L_{\alpha} + L_{\beta}}{N (N - 1)}= \nonumber \\
                                               & = & m^{\alpha, \beta} \frac{L_{\alpha} + L_{\beta}}{N (N - 1)}
\end{eqnarray}

Moreover, the average value of $ a_{ij}^{\alpha} $ over all the pairs of nodes in layer $ \alpha $ is given by:
\begin{eqnarray}
\langle a_{ij}^{\alpha} \rangle = \frac{2 L^{\alpha}}{N (N - 1)}
\end{eqnarray}
and similarly for layer $ \beta $:
\begin{eqnarray}
\langle a_{ij}^{\beta} \rangle = \frac{2 L^{\beta}}{N (N - 1)}
\end{eqnarray}
Hence, 
\begin{eqnarray}
\langle a_{ij}^{\alpha} \rangle \langle a_{ij}^{\beta} \rangle = \frac{4 L^{\alpha} L^{\beta}}{N^2 (N - 1)^2}
\end{eqnarray}
On the contrary, the expected value of multiplexity under random graph is given by:
\begin{eqnarray}
\langle m^{\alpha, \beta} \rangle & = & \frac{ 2 \sum_{i < j}^{} p^{\alpha} p^{\beta} }{L_{\alpha} + L_{\beta}} = \nonumber \\
                                  & = & \frac{N (N - 1)}{L_{\alpha} + L_{\beta}} \frac{2 L_{\alpha}}{N (N - 1)} \frac{2 L_{\beta}}{N (N - 1)} = \nonumber \\
                                  & = & \frac{1}{N (N - 1)} \frac{4 L_{\alpha} L_{\beta}}{L_{\alpha} + L_{\beta}}
\end{eqnarray}
There is therefore a direct relation between $ \langle a_{ij}^{\alpha} \rangle \langle a_{ij}^{\beta} \rangle $ and $ \langle m^{\alpha, \beta} \rangle $:
\begin{eqnarray}
\langle a_{ij}^{\alpha} \rangle \langle a_{ij}^{\beta} \rangle & = & \frac{4 L^{\alpha} L^{\beta}}{N^2 (N - 1)^2} = \nonumber \\
                                                               & = & \langle m^{\alpha, \beta} \rangle \frac{L_{\alpha} + L_{\beta}}{N (N - 1)}
\end{eqnarray}
Furthermore, we need to derive the expression for the standard deviation $ \sigma_{\alpha} $ and $ \sigma_{\beta} $:
\begin{eqnarray}
\sigma_{\alpha} & = & \sqrt{ \langle \left( a_{ij}^{\alpha} \right)^2 \rangle - \langle a_{ij}^{\alpha} \rangle^2 } = \nonumber \\
                & = & \sqrt{\langle a_{ij}^{\alpha} \rangle \left( 1 - \langle a_{ij}^{\alpha} \right) } = \nonumber \\
                & = & \sqrt{\frac{2 L_{\alpha}}{N (N - 1)} \left[ 1 - \frac{2 L_{\alpha}}{N (N - 1)} \right] }
\end{eqnarray}
and analogously for $ \beta $.
Hence, the correlation coefficient between $ a_{ij}^{\alpha} $ and $ a_{ij}^{\beta} $ is given by:
\begin{eqnarray}
Corr \{ a_{ij}^{\alpha}, a_{ij}^{\beta} \} & = & \frac{ \frac{L_{\alpha} + L_{\beta}}{N (N - 1)} m^{\alpha, \beta} - \frac{L_{\alpha} + L_{\beta}}{N (N - 1)} \langle m^{\alpha, \beta} \rangle }{ \frac{2}{N( N - 1)} \sqrt{L_{\alpha} L_{\beta} \left( 1 - \frac{2 L_{\alpha}}{N (N - 1)} \right) \left( 1 - \frac{2 L_{\alpha}}{N (N - 1)} \right)} } \nonumber \\
                                           & = & \frac{ \left( L_{\alpha} + L_{\beta} \right) \left( m^{\alpha, \beta} - \langle m^{\alpha, \beta} \rangle \right) }{ 2 \sqrt{L_{\alpha} L_{\beta} \left( 1 - \frac{2 L_{\alpha}}{N (N - 1)} \right) \left( 1 - \frac{2 L_{\alpha}}{N (N - 1)} \right)} }                                     
\end{eqnarray}
It is therefore clear that, apart from a different normalization factor (depending on $ L_{\alpha} $ and $ L_{\beta} $), our definition of binary rescaled multiplexity, when 
the random graph is considered as null model, reduces to the usual correlation coefficient~\ref{def_corr_coeff}.

However, such a property does not hold when a different reference model, such as the configuration model, is considered.

\section{Expected value for the minimum of two variables}

\subsection{Binary case}

In order to calculate (\ref{expected_bin_mul}) we need to compute the expression of the expected value of the minimum of two variables. In the unweighted case, this is easy
because it reduces to the evaluation of the expected minimum between two indipendent, binary variables. In particular, when the configuration model is considered (the extension to the
random graph is straightforward), the probability that a link exists between nodes $ i $ and $ j $ is given by the mass probability function of a Bernoulli-distributed variable: 
\begin{eqnarray}
P(a_{ij}^{\alpha}) = p_{ij}^{a_{ij}} (1 - p_{ij})^{(1 - a_{ij})}
\end{eqnarray}
Therefore, we have:
\begin{eqnarray}
\langle \min \{ a_{ij}^{\alpha}, a_{ij}^{\beta} \} \rangle_{BCM} & = & \sum_{a_{ij}^{\alpha}, a_{ij}^{\beta}}^{} \min \{ a_{ij}^{\alpha}, a_{ij}^{\beta} \} P \left(\min \{ a_{ij}^{\alpha}, a_{ij}^{\beta} \} \right) = \nonumber \\
                                                                 & = & 0 \cdotp P \left(\min \{ a_{ij}^{\alpha}, a_{ij}^{\beta} \} = 0 \right) + 1 \cdotp P \left(\min \{ a_{ij}^{\alpha}, a_{ij}^{\beta} \} = 1 \right) = \nonumber \\
                                                                 & = & P \left(\min \{ a_{ij}^{\alpha}, a_{ij}^{\beta} \} = 1 \right) = \nonumber \\
                                                                 & = & P \left(a_{ij}^{\alpha} = 1\right) P \left(a_{ij}^{\beta} = 1 \right) = \nonumber \\
                                                                 & = & p_{ij}^{\alpha} p_{ij}^{\beta}
\end{eqnarray}

Hence, the expected value of the binary multiplexity becomes:
%%%%%%%%% passaggi????
\begin{eqnarray}
\mu^{\alpha, \beta}_{BCM} = \frac{2 \sum_{i < j}^{} \left(\min \{ a_{ij}^{\alpha}, a_{ij}^{\beta} \} - p_{ij}^{\alpha} p_{ij}^{\beta}\right)}{\sum_{i < j}^{} \left(a_{ij}^{\alpha} + a_{ij}^{\beta} - 2 p_{ij}^{\alpha} p_{ij}^{\beta}\right)}
\end{eqnarray}

%%%%%%%%%%
%% z-score
In order to compute the z-score related to the multiplexity, we should evaluate also the expected value of the square of the minimum between the same two variables. Indeed, 
we have:
\begin{eqnarray}
z \left[m^{\alpha, \beta} \right] = \frac{\sum_{i < j}^{} \min \{ a_{ij}^{\alpha}, a_{ij}^{\beta}\} - \sum_{i < j}^{} \langle \min \{ a_{ij}^{\alpha}, a_{ij}^{\beta}\} \rangle}{\sigma \left[\sum_{i < j}^{} \min \{ a_{ij}^{\alpha}, a_{ij}^{\beta}\} \right]}
\end{eqnarray}
where we can evaluate the variance in the following way: 
\begin{eqnarray}
\sigma^2 \left[\min \{ a_{ij}^{\alpha}, a_{ij}^{\beta}\} \right] = \langle {\min}^2 \{ a_{ij}^{\alpha}, a_{ij}^{\beta} \} \rangle - \langle \min \{ a_{ij}^{\alpha}, a_{ij}^{\beta} \} \rangle^2 \nonumber \\
\end{eqnarray}
Exploiting again the binary character of the two indipendent variables $ a_{ij}^{\alpha} $ and $ a_{ij}^{\beta} $, the expected value of the square of the minimum becomes:
\begin{eqnarray}
\langle {\min}^2 \{ a_{ij}^{\alpha}, a_{ij}^{\beta} \} \rangle_{BCM} & = & \sum_{a_{ij}^{\alpha}, a_{ij}^{\beta}}^{} {\min}^2 \{ a_{ij}^{\alpha}, a_{ij}^{\beta} \} P\left(\min \{ a_{ij}^{\alpha}, a_{ij}^{\beta} \} \right) = \nonumber \\
                                                                     & = & 0 \cdotp P \left(\min \{ a_{ij}^{\alpha}, a_{ij}^{\beta} \} = 0 \right) + 1 \cdotp P \left(\min \{ a_{ij}^{\alpha}, a_{ij}^{\beta} \} = 1 \right) = \nonumber \\
                                                                     & = & P \left(\min \{ a_{ij}^{\alpha}, a_{ij}^{\beta} \} = 1 \right) = \nonumber \\
                                                                     & = & P\left(a_{ij}^{\alpha} = 1 \right) P \left(a_{ij}^{\beta} = 1 \right) = \nonumber \\
                                                                     & = & p_{ij}^{\alpha} p_{ij}^{\beta}
\end{eqnarray}
Therefore, the standard deviation, required in order to evaluate the z-score associated to the multiplexity, is given by:
\begin{eqnarray}
\sigma \left[\sum_{i < j}^{} \min \{ a_{ij}^{\alpha}, a_{ij}^{\beta}\} \right] = \sqrt{\sum_{i < j}^{} \left[p_{ij}^{\alpha} p_{ij}^{\beta} - \left(p_{ij}^{\alpha} p_{ij}^{\beta}\right)^2 \right]}
\end{eqnarray}

The analytical value of the z-score related to the binary multiplexity, when the configuration model is taken into account, is then:
%%%%%%%%% passaggi ?????
\begin{eqnarray}
z_{BCM} = \frac{\sum_{i < j}^{} \min \{ a_{ij}^{\alpha}, a_{ij}^{\beta}\} - \sum_{i < j}^{} p_{ij}^{\alpha}p_{ij}^{\beta}}{\sqrt{\sum_{i < j}^{} \left[p_{ij}^{\alpha}p_{ij}^{\beta} - \left(p_{ij}^{\alpha}p_{ij}^{\beta}\right)^2 \right]}}
\end{eqnarray}
Extending such results to the random graph is immediate, since everything reduces to a change in the definition of the probability of observing a link between any given pair
of nodes in each layer.

\subsection{Weighted case}

We now consider the weighted configuration model as a benchmark, but the extension to the weighted random graph is immediate.
Similarly to the binary case, the problem of finding the analytical expression for the rescaled weighted multiplexity reduces 
to the evaluation of the expected minimum between two indipendent
variables $ w_{ij}^{\alpha} $ and $ w_{ij}^{\beta} $, distributed according to a geometrical distribution:
\begin{eqnarray}
P(w_{ij}^{\alpha}) = p_{ij}^{w_{ij}^{\alpha}} (1 - p_{ij}^{\alpha})
\end{eqnarray}

In order to quantify such an expectation value, we exploit the cumulative distribution of the minimum between the considered variables:
\begin{eqnarray}
P\left(\min \{ w_{ij}^{\alpha}, w_{ij}^{\beta} \} \geq w\right) & = & P \left( w_{ij}^{\alpha} \geq w \right) P \left(w_{ij}^{\beta} \geq w \right) = \nonumber \\
                                                                & = & \left(p_{ij}^{\alpha} p_{ij}^{\beta}\right)^w
\end{eqnarray}
Thus, the expected minimum becomes:
\begin{eqnarray}
\langle \min \{ w_{ij}^{\alpha}, w_{ij}^{\beta} \} \rangle_{WCM} & = & \sum_{w'}^{} w' [ P \left(\min \{ w_{ij}^{\alpha}, w_{ij}^{\beta} \} \geq w' \right) - P \left( \min \{ w_{ij}^{\alpha}, w_{ij}^{\beta} \} \geq w' +1 \right) ] = \nonumber \\
                                                                 & = & \sum_{w'}^{} w' \left[ \left( p_{ij}^{\alpha} p_{ij}^{\beta} \right)^{w'} - \left( p_{ij}^{\alpha} p_{ij}^{\beta} \right)^{w'+1} \right] = \nonumber \\
                                                                 & = & \frac{p_{ij}^{\alpha} p_{ij}^{\beta}}{1 - p_{ij}^{\alpha} p_{ij}^{\beta}}
\label{exp_min}
\end{eqnarray}
which leads to following expression for the weighted multiplexity under weighted configuration model:
%%%%%%%%% passaggi????
\begin{eqnarray}
\mu^{\alpha, \beta}_{WCM} = \frac{2 \sum_{i < j}^{} \left( \min \{ w_{ij}^{\alpha}, w_{ij}^{\beta} \} - \frac{p_{ij}^{\alpha} p_{ij}^{\beta}}{1 - p_{ij}^{\alpha} p_{ij}^{\beta}} \right)}{\sum_{i < j}^{} \left( w_{ij}^{\alpha} + w_{ij}^{\beta} - 2 \frac{p_{ij}^{\alpha} p_{ij}^{\beta}}{1 - p_{ij}^{\alpha} p_{ij}^{\beta}} \right) }
\end{eqnarray}

%%%%%%%%%%%
%% z-score
About the z-score referred to the weighted multiplexity, we can define it in the usual way:
\begin{eqnarray}
z \left[ m^{\alpha, \beta} \right] = \frac{\sum_{i < j}^{} \min \{ w_{ij}^{\alpha}, w_{ij}^{\beta}\} - \sum_{i < j}^{} \langle \min \{ w_{ij}^{\alpha}, w_{ij}^{\beta}\} \rangle}{\sigma \left[ \sum_{i < j}^{} \min \{ w_{ij}^{\alpha}, w_{ij}^{\beta}\} \right]}
\end{eqnarray}
Since:
\begin{eqnarray}
\sigma^2 \left[ \min \{ w_{ij}^{\alpha}, w_{ij}^{\beta}\} \right] = \langle {\min}^2 \{ w_{ij}^{\alpha}, w_{ij}^{\beta} \} \rangle - \langle \min \{ w_{ij}^{\alpha}, w_{ij}^{\beta} \} \rangle^2 \nonumber \\
\end{eqnarray}
we just have to compute the analytical expression for the expected value of the square of minimum bewteeen $ w_{ij}^{\alpha} $ and $ w_{ij}^{\beta} $. Then, following the same
procedure adopted for (\ref{exp_min}) we find:
\begin{eqnarray}
\langle {\min}^2 \{ w_{ij}^{\alpha}, w_{ij}^{\beta} \} \rangle_{WCM} & = & \sum_{w'}^{} \left( w' \right)^2 [ P \left( \min \{ w_{ij}^{\alpha}, w_{ij}^{\beta} \} \geq w' \right) - P \left( \min \{ w_{ij}^{\alpha}, w_{ij}^{\beta} \} \geq w'+1 \right) ]  = \nonumber \\
                                                                     & = & \sum_{w'}^{} \left( w' \right)^2 \left[ \left( p_{ij}^{\alpha} p_{ij}^{\beta} \right)^{w'} - \left( p_{ij}^{\alpha} p_{ij}^{\beta}\right)^{w'+1}\right] = \nonumber \\
                                                                     & = & \frac{p_{ij}^{\alpha} p_{ij}^{\beta} + \left( p_{ij}^{\alpha} p_{ij}^{\beta} \right)^2}{\left(1 - p_{ij}^{\alpha} p_{ij}^{\beta} \right)^2}
\end{eqnarray}
and therefore the standard deviation is:
\begin{eqnarray}
\sigma \left[\sum_{i < j}^{} \min \{ w_{ij}^{\alpha}, w_{ij}^{\beta}\}\right] = \sqrt{\sum_{i < j}^{} \left[ \frac{p_{ij}^{\alpha} p_{ij}^{\beta} + \left(p_{ij}^{\alpha} p_{ij}^{\beta}\right)^2}{\left(1 - p_{ij}^{\alpha} p_{ij}^{\beta}\right)^2} - \frac{\left(p_{ij}^{\alpha} p_{ij}^{\beta}\right)^2}{\left(1 - p_{ij}^{\alpha} p_{ij}^{\beta}\right)^2} \right] }
\end{eqnarray}
Finally, the z-score associated to the weighted multiplexity under configuration model is therefore given by:
\begin{eqnarray}
z_{WCM} = \frac{\sum_{i < j}^{} \min \{ w_{ij}^{\alpha}, w_{ij}^{\beta}\} - \sum_{i < j}^{} \frac{p_{ij}^{\alpha}p_{ij}^{\beta}}{1 - p_{ij}^{\alpha}p_{ij}^{\beta}}}{\sqrt{\sum_{i < j}^{} \frac{p_{ij}^{\alpha}p_{ij}^{\beta}}{\left(1 - p_{ij}^{\alpha}p_{ij}^{\beta}\right)^2}}}
\end{eqnarray}

%%%%%%%%%%%%%%%%%%%
\section{International Trade Network and European Airport Network: list of layers}

In Table \ref{tab:commodities} we report the list of commodities, according to the standard international classification HS1996~\cite{Comtrade}, traded in 2011. In Table \ref{tab:airlines}, instead, we report the list of airlines present in the European Airport
dataset we considered in the main text.

\newpage

\begin{multicols*}{2}
\TrickSupertabularIntoMulticols
\begin{supertabular}{|c | >{\centering\arraybackslash}p{5cm} | c|}
\hline
Code & Commodity & Class\\
\hline
01 & Live animals & \mycirc[purple]\\ 
%\hline
02 & Meat and edible meat offal & \mycirc[purple]\\ 
%\hline
03 & Fish, crustaceans and acquatic invertebrates & \mycirc[purple]\\
%\hline
04 & Dairy produce; birs eggs; honey and other edible animal products & \mycirc[purple]\\
%\hline
05 & Other products of animal origin & \mycirc[purple]\\
%\hline
06 & Live trees, plants; bulbs, roots; cut flowers and ornamental foliage tea and spices & \mycirc[red]\\
%\hline
07 & Edible vegetables and certain roots and tubers & \mycirc[red]\\
%\hline
08 & Edible fruit and nuts; citrus fruit or melon peel & \mycirc[red]\\
%\hline
09 & Coffee, tea, mate and spices & \mycirc[red]\\
%\hline
10 & Cereals & \mycirc[red]\\
%\hline
11 & Milling products; malt; starch; inulin; wheat gluten & \mycirc[red]\\ 
%\hline
12 & Oil seeds and oleaginous fruits; miscellaneous grains, seeds and fruit; industrial or medicinal plants; straw and fodder & \mycirc[red]\\ 
%\hline
13 & Lac; gums, resins and other vegetable sap and extracts & \mycirc[red]\\
%\hline
14 & Vegetable plaiting materials and other vegetable products & \mycirc[red]\\
%\hline
15 & Animal, vegetable fats and oils, cleavage products, etc. & \mycirc[red]\\
%\hline
16 & Edible preparations of meat, fish, crustaceans, mollusks or other aquatic invertebrates & \mycirc[red]\\
%\hline
17 & Sugars and sugar confectionary & \mycirc[red]\\
%\hline
18 & Cocoa and cocoa preparations & \mycirc[red]\\
%\hline
19 & Preparations of cereals, flour, starch or milk; bakers wares & \mycirc[red]\\
%\hline
20 & Preparations of vegetables, fruit, nuts or other plant parts & \mycirc[red]\\
%\hline
21 & Miscellaneous edible preparations & \mycirc[red]\\ 
%\hline
22 & Beverages, spirits and vinegar & \mycirc[red]\\ 
%\hline
23 & Food industry residues and waste; prepared animal feed & \mycirc[red]\\
%\hline
24 & Tobacco and manufactured tobacco substitutes & \mycirc[red]\\
%\hline
25 & Salt; sulfur; earth and stone; lime and cement plaster & \mycirc[blue]\\
%\hline
26 & Ores, slag and ash & \mycirc[blue]\\
%\hline
27 & Mineral fuels, mineral oils and products of their distillation; bitumin substances; mineral wax & \mycirc[blue]\\
%\hline
28 & Inorganic chemicals; organic or inorganic compounds of precious metals, of rare-earth metals, of radioactive elements or of isotopes & \mycirc[blue] \\
%\hline
29 & Organic chemicals & \mycirc[blue]\\
%\hline
30 & Pharmaceutcal products & \mycirc[blue]\\
%\hline
31 & Fertilizers & \mycirc[blue]\\ 
%\hline
32 & Tanning or dyeing extracts; tannins and derivatives; dyes, pigments and coloring matter; paint and varnish; putty and other mastics; inks & \mycirc[blue]\\ 
%\hline
33 & Essential oils and resinoids; perfumery, cosmetic or toilet preparations & \mycirc[blue]\\
%\hline
34 & Soap; waxes; polish; candles; modeling pastes; dental preparations with basic of plaster & \mycirc[blue]\\
%\hline
35 & Albuminoidal substances; modified starch; glues; enzymes & \mycirc[blue]\\
%\hline
36 & Explosives; pyrotechnic products; matches; pyrophoric alloys; certain combustible preparations & \mycirc[cyan]\\
%\hline
37 & Photographic or cinematographic goods & \mycirc[cyan]\\
%\hline
38 & Miscellaneous chemical products & \mycirc[cyan]\\
%\hline
39 & Plastics and articles thereof & \mycirc[cyan]\\
%\hline
40 & Rubber and articles thereof& \mycirc[cyan] \\
%\hline
41 & Raw hides and skins (other than furskins) and leather & \mycirc[brown]\\ 
%\hline
42 & Leather articles; saddlery and harness; travel goods, handbags and similar; articles of animal gut (not silkworm gut) & \mycirc[brown]\\ 
%\hline
43 & Furskins and artificial fur; manufactures thereof & \mycirc[brown]\\
%\hline
44 & Wood and articles of wood; wood charcoal & \mycirc[brown]\\
%\hline
45 & Cork and articles of cork & \mycirc[brown]\\
%\hline
46 & Manufactures of straw, esparto or other plaiting materials; basketware and wickerwork & \mycirc[brown]\\
%\hline
47 & Pulp of wood or of other fibrous cellulosic material; waste and scrap of paper and paperboard & \mycirc[brown]\\
%\hline
48 & Paper and paperboard and articles thereof; paper pulp articles & \mycirc[brown]\\
%\hline
49 & Printed books, newspapers, pictures and other products of printing industry; manuscripts, typescripts & \mycirc[brown]\\
%\hline
50 & Silk, including yarns and woven fabric thereof & \mycirc[orange]\\
%\hline
51 & Wool and animal hair, including yarn and woven fabric & \mycirc[orange]\\ 
%\hline
52 & Cotton, including yarn and woven fabric thereof & \mycirc[orange]\\ 
%\hline
53 & Other vegetable textile fibers; paper yarn and woven fabrics of paper yarn & \mycirc[orange]\\
%\hline
54 & Manmade filaments, including yarns and woven fabrics & \mycirc[orange]\\
%\hline
55 & Manmade staple fibers, including yarns and woven fabrics & \mycirc[orange]\\
%\hline
56 & Wadding, felt and nonwovens; special yarns; twine, cordage, ropes and cables and article thereof & \mycirc[orange]\\
%\hline
57 & Carpets and other textile floor coverings & \mycirc[orange]\\
%\hline
58 & Special woven fabrics; tufted textile fabrics; lace; tapestries; trimmings; embroidery & \mycirc[yellow]\\
%\hline
59 & Impregnated, coated, covered or laminated textile fabrics; textile articles for industrial use & \mycirc[yellow]\\
%\hline
60 & Knitted or crocheted fabrics & \mycirc[yellow]\\
%\hline
61 & Apparel articles and accessories, knitted or crocheted & \mycirc[yellow]\\ 
%\hline
62 & Apparel articles and accessories, not knitted or crocheted & \mycirc[yellow]\\ 
%\hline
63 & Other textile articles; needlecraft sets; worn clothing and worn textile articles; rags & \mycirc[yellow]\\
%\hline
64 & Footwear, gaiters and the like and parts thereof & \mycirc[yellow]\\
%\hline
65 & Headgear and parts thereof & \mycirc[yellow]\\
%\hline
66 & Umbrellas, walking sticks, seat sticks, riding crops, whips, and parts thereof & \mycirc[yellow]\\
%\hline
67 & Prepared feathers, down and articles thereof; artificial flowers; articles of human hair & \mycirc[yellow]\\
%\hline
68 & Articles of stone, plaster, cement, asbestos, mica or similar materials & \mycirc[green]\\
%\hline
69 & Ceramic products & \mycirc[green]\\
%\hline
70 & Glass and glassware & \mycirc[green]\\
%\hline
71 & Pearls, precious stones, metals, coins, etc. & \mycirc[green]\\ 
%\hline
72 & Iron and steel & \mycirc[green]\\ 
%\hline
73 & Articles of iron and steel & \mycirc[green]\\
%\hline
74 & Copper and articles thereof & \mycirc[green]\\
%\hline
75 & Nickel and articles thereof & \mycirc[green]\\
%\hline
76 & Aluminum and articles thereof & \mycirc[green]\\
%\hline
77 & Lead and articles thereof & \mycirc[green]\\
%\hline
78 & Zinc and articles thereof & \mycirc[green]\\
%\hline
79 & Tin and articles thereof & \mycirc[green]\\
%\hline
80 & Other base metals; cermets; articles thereof & \mycirc[green]\\
%\hline
81 & Tools, implements, cutlery, spoons and forks of base metal and parts thereof & \mycirc[green]\\ 
%\hline
82 & Miscellaneous articles of base metal & \mycirc[green]\\ 
%\hline
83 & Nuclear reactors, boilers, machinery and mechanical appliances; parts thereof & \mycirc[darkgray]\\
%\hline
84 & Electric machinery, equipment and parts; sound equipment; television equipment & \mycirc[darkgray]\\
%\hline
85 & Railway or tramway; locomotives, rolling stock, track fixtures and parts thereof; mechanical and electromechanical traffic signal equipment & \mycirc[darkgray]\\
%\hline
86 & Vehicles (not railway, tramway, rolling stock); parts and accessories & \mycirc[darkgray]\\
%\hline
87 & Aircraft, spacecraft, and parts thereof & \mycirc[darkgray]\\
%\hline
88 & Ships, boats and floating structures & \mycirc[darkgray]\\
%\hline
89 & Optical, photographic, cinematographic, measuring, checking, precision, medical or surgical instruments/apparatus; parts and accessories & \mycirc[black]\\
%\hline
90 & Clocks and watches and parts thereof & \mycirc[black]\\
%\hline
91 & Musical instruments; parts and accessories thereof & \mycirc[black]\\ 
%\hline
92 & Arms and ammunition, parts and accessories thereof & \mycirc[black]\\ 
%\hline
93 & Furniture; bedding, mattresses, cushions, etc.; other lamps and light fitting, illuminated signs and nameplates, prefabricate buildings & \mycirc[black]\\
%\hline
94 & Toys, games and sports equipment; parts and accessories & \mycirc[black]\\
%\hline
95 & Miscellaneous manufactured articles & \mycirc[black]\\
%\hline
96 & Works of art, collectors pieces and antiques & \mycirc[black]\\
\hline
\end{supertabular}
\captionof{table}{List of commodities, according to the standard international classification HS1996, and associated codes. In the third column 
we divide such commodities in classes of similar traded items, each of them being represented by a different colored circle.
Colors are the same as reported in the dendrogram in Figure~\ref{fig:dendro_BACI}. }
\label{tab:commodities}
\end{multicols*}

\newpage 

\begin{multicols*}{2}
\TrickSupertabularIntoMulticols
\begin{supertabular}{|c|c|}
\hline
Code & Airline \\
\hline
1 & NextJet \\
2 & Fly 6ix \\
3 & Air Berlin \\
4 & Air France \\
5 & Finnair \\
6 & Alitalia \\
7 & British Airways \\
8 & Air Sicilia \\
9 & Air Baltic \\
10 & Air China \\
11 & DAT Danish Air Transport \\
12 & Norwegian Air Shuttle \\
13 & Aer Lingus \\
14 & Icelandair \\
15 & Air Greenland \\
16 & Niki \\
17 & Hainan Airlines \\
18 & Air Bosna \\
19 & Adria Airways \\
20 & Jat Airways \\
21 & KLM Royal Dutch Airlines \\
22 & Luxair \\
23 & Lufthansa \\
24 & LOT Polish Airlines \\
25 & Swiss International Air Lines \\
26 & Czech Airlines \\
27 & Croatia Airlines \\
28 & Estonian Air \\
29 & Pegasus Airlines \\
30 & Pakistan International Airlines \\
31 & Atlantic Airways \\
32 & Scandinavian Airlines System \\
33 & Brussels Airlines \\
34 & Singapore Airlines \\
35 & Aeroflot Russian Airlines \\
36 & Turkish Airlines \\
37 & TAP Portugal \\
38 & easyJet \\
38 & United Feeder Service \\
39 & Vueling Airlines \\
40 & Wider\o{}e \\
42 & Azerbaijan Airlines \\
43 & Atlasjet \\
44 & AirOnix \\
45 & Ukraine International Airlines \\
46 & Golden Air \\
47 & Airlinair \\
48 & Livingston \\
49 & Aegean Airlines \\
50 & Belavia Belarusian Airlines \\
51 & El Al Israel Airlines \\
52 & Wizz Air \\
53 & Ryanair \\
54 & Gazpromavia \\
55 & Ciel Canadien \\
56 & Germanwings \\
57 & Flybe \\
58 & Condor Flugdienst \\
59 & Bulgaria Air \\
60 & Transavia Holland \\
61 & Jet2.com \\
62 & Transavia France \\
63 & Transaero Airlines \\
64 & Air Europa \\
65 & Maastricht Airlines \\
66 & Monarch Airlines \\
67 & Air Bourbon \\
68 & Ural Airlines \\
69 & Binter Canarias \\
70 & Germania \\
71 & Big Sky Airlines \\
72 & Olympic Airlines \\
73 & S7 Airlines \\
74 & Cargoitalia \\
75 & Aircompany Yakutia \\
76 & NordStar Airlines \\
77 & Eastern Airways \\
78 & Aerocondor \\
79 & BRA-Transportes Aereos \\
80 & Rossiya-Russian Airlines \\
81 & Vladivostok Air \\
82 & Tarom \\
83 & Air Moldova \\
84 & Aerolineas Argentinas \\
85 & Iberia Airlines \\
86 & LAN Airlines \\
87 & China Eastern Airlines \\
88 & Air Iceland \\
89 & Twin Jet \\
90 & Cyprus Airways \\
91 & Air Malta"  \\
92 & Starling Airlines Spain \\
93 & Air Service \\
94 & Virgin Atlantic Airways \\
95 & Wizz Air Ukraine \\
96 & SunExpress \\
97 & Qatar Airways \\
98 & Montenegro Airlines \\
99 & Nationwide Airlines \\
100 & Carpatair \\
101 & Flybaboo \\
102 & Egyptair \\
103 & SATA Air Acores \\
104 & Meridiana \\
105 & Balkan Bulgarian Airlines \\
106 & BAL Bashkirian Airlines \\
107 & Star1 Airlines \\
108 & Air One \\                                                                                                         
109 & Blue Panorama Airlines \\
110 & Ethiopian Airlines \\
111 & Korean Air \\
112 & Teamline Air \\
113 & Oman Air \\
114 & Corse-Mediterranee \\ 
115 & East African \\
116 & Orenburg Airlines \\
117 & Air Senegal International \\
118 & Cielos Airlines \\
119 & SATA International \\
120 & TUIfly \\
121 & Israir \\
122 & Eurolot \\
123 & Kuwait Airways \\
124 & Travel Service \\
125 & Onur Air \\
126 & Malm\"{o} Aviation \\
127 & Hex'Air \\    %%%%%%%%%%%%%%%%
128 & Aigle Azur \\
129 & Aeroflot-Nord \\
130 & Tulip Air \\
131 & Moskovia Airlines \\
132 & Intersky \\
133 & Belair Airlines \\
134 & Royal Jordanian \\
135 & Motor Sich \\
136 & TransHolding System \\
137 & SmartLynx Airlines \\
138 & China Airlines \\
139 & Air Dolomiti \\
140 & Polet Airlines \\
141 & Etihad Airways \\
142 & Air Transat \\
143 & Saratov Aviation Division \\
144 & LTU International \\
145 & Atlantis European Airways \\
146 & Tatarstan Airlines \\
147 & Austrian Airlines \\
148 & Georgian Airways \\
149 & UTair-Express \\
150 & Four Star Aviation / Four Star Cargo \\
151 & Tropic Air \\
152 & XL Airways France \\
153 & Ozark Air Lines \\
153 & Asiana Airlines \\
154 & Air Europe \\                       
155 & SkyWork Airlines \\                      
156 & Danube Wings \\
157 & Air Kazakhstan \\
158 & Hahn Air \\
159 & Alaska Central Express \\
160 & IzAvia \\
161 & Air Armenia \\
162 & Arkia Israel Airlines \\
163 & Sat Airlines \\
164 & North Flying \\                           
165 & JobAir \\
166 & Emirates \\
167 & Yellowtail \\
168 & Royal Air Maroc \\
169 & MIAT Mongolian Airlines \\
170 & Uzbekistan Airways \\
171 & Ariana Afghan Airlines \\
\hline 
\end{supertabular}
\captionof{table}{List of airlines operating in Europe, as provided by the OpenFlight database. }
\label{tab:airlines}
\end{multicols*}

\end{document}